\documentclass[twocolumn,showpacs,preprintnumbers,amsmath,amssymb]{revtex4}

\usepackage{graphicx}
\usepackage{dcolumn}
\usepackage{bm}

\begin{document}

\preprint{APS}

\title{ Current and noise expressions for radio-frequency single-electron transistors}
\author{ Jung Hyun Oh,$^1$}
\email{jungoh@iquips.uos.ac.kr}
\author{D. Ahn,$^1$ and S. W. Hwang$^{1,2}$}

\affiliation{$^1$ Institute of Quantum Information Processing and
Systems, University of Seoul, 90 Jeonnong, Tongdaemoon-ku, Seoul
130-743, Korea\\
$^2$ Department of Electronics Engineering, Korea University, Anam,
Sungbuk-ku, Seoul 136-075, Korea}

\date{\today}

\begin{abstract}
We derive self-consistent expressions of current and noise for single-electron transistors
driven by time-dependent perturbations.
We take into account effects of the electrical environment, higher-order co-tunneling, and 
time-dependent perturbations under the two-charged state approximation
using the Schwinger-Kedysh approach combined with the generating functional technique.
For a given generating functional, we derive exact expressions for tunneling currents and noises
and present the forms in terms of transport coefficients.
It is also shown that in the adiabatic limit our results encompass previous formulas.
In order to reveal effects missing in static cases, we apply the derived results
to simulate realized radio-frequency single-electron transistor.
It is found that photon-assisted tunneling  affects largely the performance of
the single-electron transistor by enhancing both responses to gate charges and current noises.
On various tunneling resistances and frequencies of microwaves,  the dependence of
the charge sensitivity is also discussed.
\end{abstract}


\maketitle

\section{Introduction}

For a metallic island small enough to give large charging energy exceeding the temperature,
quantum transport through it shows remarkable features
due to strongly correlated electrons.
A single-electron transistor (SET) is popular geometry of studying 
transport through the metallic island, in which the island is coupled to
two large reservoirs (source and drain) via tunneling junctions and 
to another reservoir capacitively (gate).\cite{houton,averin,ingold,kouwen,schon} 
The most important principle of operating SET is the Coulomb-blockade effects in
the island.
In an usual situation particles cannot tunnel into the island due to the Coulomb energy.
However, when the Coulomb energy is reduced by a gate bias, transport through the island
is possible even with  a small bias between the source and drain.
Consequently, tunneling currents show a series of peaks, called the Coulomb-blockade peaks,
as a function of gate voltage.

From fundamental and applied point of views, the shape of Coulomb-blockade peaks
has been attracted much attention and widely studied in static bias conditions.\cite{been}
The appearance of the Coulomb-blockade peaks depends on two conditions.
The first is  much larger charging energy than the temperature to blockade thermal excitation.
The second is  thick tunneling barriers guaranteeing a large dwelling time
for electrons to resolve the charging energy in the island.
Then, from the energy uncertainty principle, the latter is approximately fulfilled
under the condition that  parallel resistance $R_T$ of the barriers is much larger than the 
resistance quantum $R_K=h/e^2$, i.e., $\alpha_0=R_K/(4 \pi^2 R_T)\ll 1$.
It is now well known that for large tunneling resistance $R_T$,
transport in a SET is achieved by the sequences of uncorrelated tunneling processes and
the smearing of peaks is dominated by the temperature.
However,
for relatively small tunneling resistance ($\alpha_0\leq 1$) or very low temperature,
additional quantum processes such as higher-order co-tunneling contribute the peaks and
renormalizes even its positions as well as
associated quantum fluctuation is responsible for the broadening of the peaks.\cite{schon,schoeller}
It is also found that these higher-order co-tunneling effects are manifested to
noises of the system.\cite{averin1,sukhor,utsumi}
On the other hand, in concerning about applications of SETs to an electrometer,
a slope of a Coulomb-blockade peak with respect to gate voltages  is an important
factor.  Since the distance between peaks is equal to the change of an elementary
charge on the gate, a large slope of the peaks means
high sensitivity to a fraction of the charges.
So, it is widely believed that a SET is a prime candidate for reading out
the final state of a qubit in a solid-state quantum computer.\cite{makhlin,kane,aassime,johan}

New theoretical interests in transport properties through the strongly correlated systems
are emerging together with the experimental success in driving them by
microwaves (radio-frequency waves), which are called radio-frequency
single-electron transistors (rf SETs).\cite{schoel}
In such a system, microwaves are delivered via a $LC$-resonant circuit
to excite particles to overcome the Coulomb energy.
As a consequence, SETs can operate in a high-frequency domain and practically
provides advantage of a large bandwidth as an electrometer,
which allows to measure the rapid variation
of gate charges.\cite{schoel,aassime,fujisawa,cheong,buehler}
Theoretically, the interplay of electronic transport and excitations by microwaves
is a particular interest because high-frequency perturbations are expected to yield
a new non-equilibrium situation resulted from additional phase variation in energy states.\cite{tien}
Such a time-dependent situation is usually divided into classical and quantum regimes.
In the classical regime (or adiabatic regime) energies excited by time-dependent perturbation appear
to be continuous while in the quantum regime (we will also refer to this as non-adiabatic regime)
discrete photon energies become observable and 
particles can emit or absorb photons when they tunnel from an initial state on one side
of the barrier to a final state on the opposite side,
called as photon-assisted tunneling.\cite{kouwen1,bruder,staff,oh}

So, in order to understand transport properties of rf SETs,
one may need generic theoretical considerations including
higher-order co-tunneling processes as well as sequential tunneling, even
in the quantum regime of time-dependent perturbations.
Actually, according to the recent experiments of rf SETs,\cite{aassime,schoel,fujisawa,cheong,buehler}
tunneling resistances range from $\alpha_0=10^{-5}$ to $3\times 10^{-2}$, implying
feasible co-tunneling processes for large values of $\alpha_0$.
Frequencies of microwaves were used from 0.3 to 1.7 GHz,
which correspond to several $\mu eV$ of photon energies,
comparable or larger than  thermal energy in the experiments.
For these values of frequencies, it has been not known for the system to be
driven in the classical or quantum regimes of time-dependent perturbations.
So, in general it is necessary to solve problems in the quantum regime
for more rigorous understanding of transport in rf SETs.

Additionally, tunneling in a rf SET may be dissipative
due to a $LC$ resonant circuit.
Since a microwave is delivered via a coaxial cable with $50\Omega$ impedance much smaller
than the resistance quantum,
effects of the electrical environment is usually ignored.
This is the case for a $LC$ resonant circuit with a low quality factor, however,
with a high-quality factor, energy states of the electrical environment become long-lived
because it becomes similar to a simple-harmonic oscillator.\cite{devoret,grabert,ingold,schon,oh1}
Then, during tunneling, particles may emit or absorb energy quanta equal to
resonant energy of the harmonic oscillator to the environment.

In this work, we develop the formalism that is capable of treating all
above theoretical considerations;
the effects of higher-order co-tunneling, non-adiabatic time-dependent perturbations,
and the electrical environments on operations of rf SETs.
Our work is a generalization of several earlier works which address the effects
partially, neglecting time-dependent perturbation,\cite{utsumi}
higher-order co-tunneling,\cite{bruder,oh} and electron-electron interaction.\cite{jauho,camalet}
However, since we use a two-charged-state model in a metallic island
assuming large charging energy, along this aspect, our work
is more restricted than Ref. \cite{bruder,schoeller,oh}.
In solving the problem, we use the Schwinger-Keldysh approach combined with
a generating functional\cite{utsumi,negele,chou,babi}
where pseudo-spins of two-charged states are treated with
the drone-fermion mapping.
Since this approach includes any higher order moment of diagrams systematically,
it is one of the well-suited methods for transport through strongly correlated system
as indicated in Ref. \cite{utsumi}.
From a generating functional summed diagrammatically, observables of the system
are obtained by functional derivatives with respect to external perturbations.
This is another advantageous point of this approach because higher-order moments
such as noise are easily calculated and expressions for observables are consistent
with each other in a sense that they are derived from the same order of diagrams.
We express the electrical environment in terms of infinite number of driven
harmonic oscillator following Caldeira and Legget\cite{caldeira} where external alternating
voltages are treated with classical fields.
Based on the unitary transformation which leaves the electrical environment 
in a stationary situation, we incorporate equilibrium fluctuation of the environment into 
the generating functional and derive environment- and time-dependent self-energies
by counting dominated diagrams.
Results for currents and noises are expressed in terms of transport coefficients.
In cases of time-dependent perturbations,
due to the displacement component currents are found to depend on
an additional transport coefficient, leading to a generalization of
the Landauer formula\cite{landauer} and noises also has its contributions.

The paper is organized as follows. 
We first describe the model of calculations in Section II. 
By expressing the dissipative environment in terms of driven harmonic oscillators,
we give the Hamiltonian depending on tunneling currents.
In Section III, we calculate an approximated generating functional based on the Schwinger-Keldysh
approach and discuss several approximations in deriving it.
For a given generating functional we show exact expressions for currents and noises in Section IV and
rewrite them in terms of transport coefficients.
In Section V, we use our formalism to simulate a rf SET
numerically and emphasize different points from static calculations.
Finally, we summarize our results in Section VI.

\section{Model of calculations}

\subsection{Hamiltonian}
\begin{figure}
\centering
\includegraphics[width=0.4\textwidth]{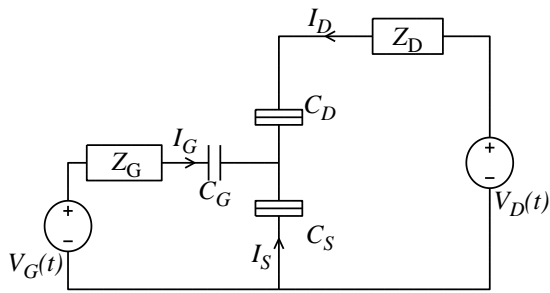}
\caption{
A typical drawing of the single-electron transistor is shown where 
time-dependent voltages are applied to a quantum dot
via possible dissipative elements connected to drain and gate electrodes, respectively.
\label{setFig}
}
\label{fig1}
\end{figure}

To formulate the problem of  a rf SET,
we begin with general circuital geometry of Fig. \ref{setFig} where time-dependent
external sources, $V_D(t)$ and $V_G(t)$, are supplied
via dissipative elements of impedances $Z_D(\omega)$
and $Z_G(\omega)$.
In this section we do not specify detailed forms
for $Z_D(\omega)$ and $Z_G(\omega)$ bearing in mind 
the application of our formalism to other systems concerning
effects of dissipative environments.\cite{rimberg,mason,penttila}
As  a typical model of a SET, a small island is coupled via tunneling barriers
to two leads, source and drain, and also capacitively to source, drain, and gates
with capacitances of $C_S$, $C_D$, and $C_G$, respectively.
We assume that the small island
is a metallic one, i.e., there are many energy levels with negligibly small
level spacing and also many particles occupied to them.
In such a metallic island, one can treat a excess charge of $Q$ confined in it
as a independent variable from those of quasiparticles in a good approximation,
and usually expresses  its Hamiltonian as,\cite{bruder,schoel,oh,shnirman}
\begin{eqnarray}
{\cal H}_{I}  = \sum_{nk} \epsilon_{kI}a_{nkI}^\dagger a_{nkI}+
\frac{Q^2}{2 C_\Sigma}
\end{eqnarray}
where 
$a_{nkI}$($a_{nkI}^\dagger$) are the annihilation (creation) operators for quasiparticles
with energy $\epsilon_{kI}$ in the island and
the index $n$ describes the transverse channels including spin.
The second term is a Coulomb-blockade model of the electron-electron interaction with
$C_{\Sigma}=C_D+C_S+C_G$.
Further simplification of the Coulomb interaction term can be made
if one uses a two-state model for excess charges.
Assuming the small island enough for charging energy $Q^2/2 C_\Sigma$
to be the largest energy scale in the problem,
it is sufficient to consider two number of charged states,
say, $\mid 0\rangle$ and $\mid 1\rangle$.
Then, the charge operator $Q$ becomes $Q=e\mid\! 1\rangle \langle 1\!\mid$ and
satisfies $Q^2=Q$ (throughout the work $e$ is the proton charge).
If we adopt spinor notation, then the Hamiltonian is further written as,
\begin{eqnarray}
{\cal H}_{I}  = \sum_{nk} \epsilon_{kI}a_{nkI}^\dagger a_{nkI}
+\Delta_0\frac{\sigma_z+1}{2}
\label{HI}
\end{eqnarray}
where $\sigma_z$ is the effective spin-1/2 operator and
$\Delta_0 = E_C(1-2 q_0/e)$ is the energy difference between the two charge
states together with the charging energy $E_C= e^2/2 C_\Sigma$.
Here, we anticipate $\Delta_0$ which depends on a static component of
a gate voltage $V_G^0$ through a charge of $q_0=C_G V_G^0$.

\begin{figure}
\centering
\includegraphics[width=0.4\textwidth]{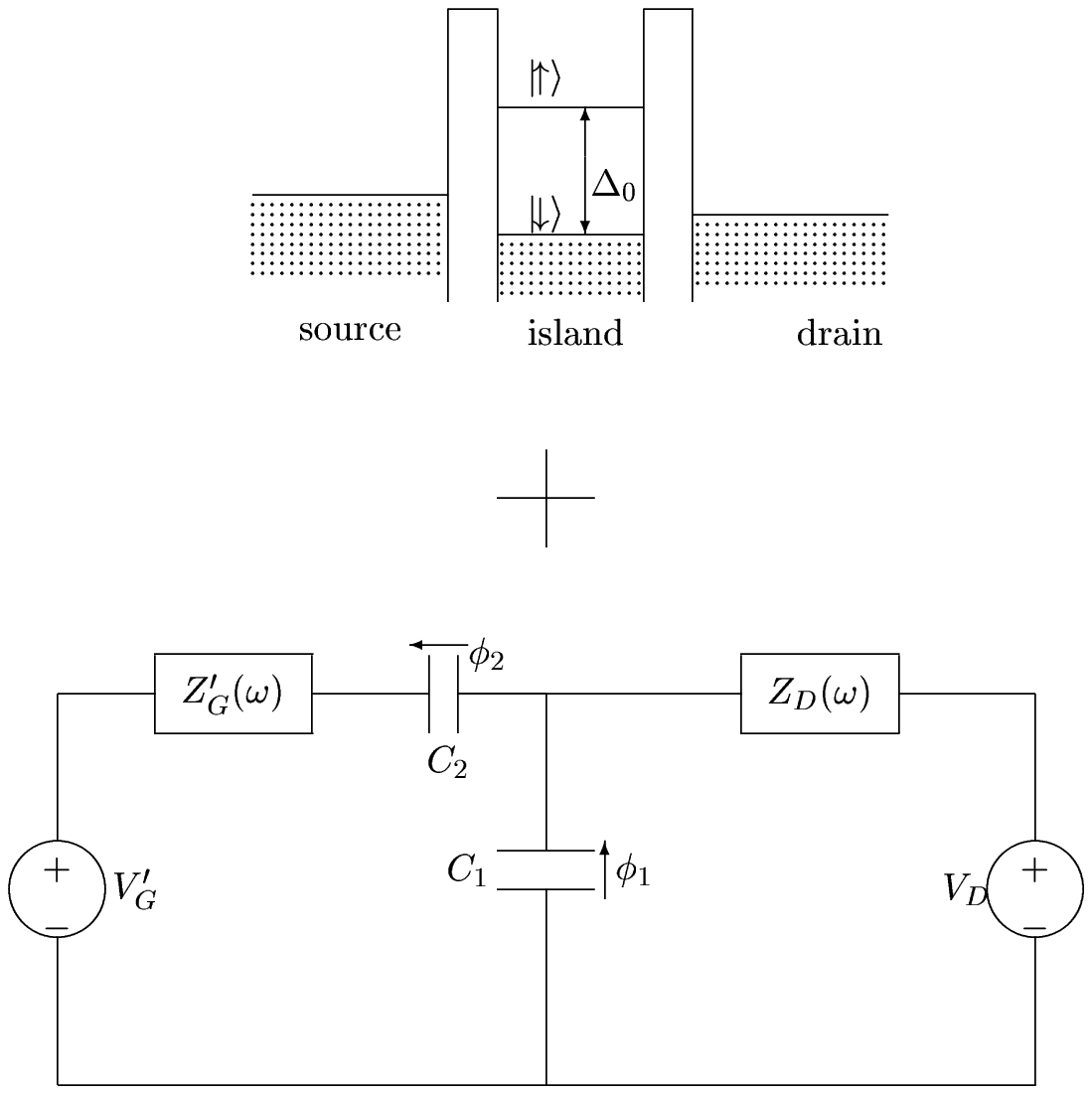}
\caption{
Microscopic and macroscopic parts of the single-electron transistor in Fig. \ref{fig1} are shown.
Here, we define $Z_G'(\omega) = Z_G(\omega)(C_D+C_S)^2/C_D^2$ and $V_G'(t)= V_G(t) (C_D+C_S)/C_D$.
}
\label{unpartFig}
\end{figure}

As for the remaining parts of the system, the Hamiltonian can be found 
by separating it into terms depending on macroscopic and microscopic variables
in a similar manner to Ref. \cite{oh}.
Then, the total Hamiltonian may be written as the sum of
the unperturbed part and the tunneling part ${\cal H}'_T$;
\begin{eqnarray}
{\cal H} ={\cal H}_I+{\cal H}_{lead}
+{\cal H}_{RLC}(V_D,V_G) + {\cal H}'_T.
\label{totH}
\end{eqnarray}
Here, ${\cal H}_I$, ${\cal H}_{lead}$, and ${\cal H}_{RLC}(V_D,V_G)$ represent
the unperturbed part of our system, which are shown in Fig. \ref{unpartFig}.
The Hamiltonian ${\cal H}_{lead}=\sum_{nk, \ell=S,D} \epsilon_{k\ell}a_{nk\ell}^\dagger a_{nk\ell}$
describes non-interacting electrons in the source and drain with
their annihilation ($a_{nk\ell}$) and creation ($a^\dagger_{nk\ell})$ operators while
the Hamiltonian ${\cal H}_{RLC}(V_D,V_G)$ governs 
the electrical environment which corresponds to a lumped circuit of Fig. \ref{unpartFig}.
Actually, the lumped circuit is designed to exhibit the same dynamical behavior as that of Fig. 1 
if it were not for tunneling and thus tunneling barriers work as just capacitances.
For this, dynamical variables in the two circuit are related to each other as,
\begin{eqnarray}
\left(
    \begin{array}{c}
Q_1\\
Q_2\\
Q  \\
     \end{array}
\right)
 &=& 
\left(
    \begin{array}{ccc}
 \frac{C_1}{C_D} & -\frac{C_1}{C_S} &  0        \\
-\frac{C_2}{C_D} & -\frac{C_2}{C_D} & \frac{C_D}{C_\Sigma}     \\
 -1      &  -1      & -1        \\
     \end{array}
\right)
\left(
    \begin{array}{c}
Q_D\\
Q_S\\
Q_G\\
     \end{array}
\right),
\label{Q}
\\
\left(
    \begin{array}{c}
\phi_D\\
\phi_S\\
     \end{array}
\right)
&=&
{\bf m }
\left(
    \begin{array}{c}
\phi_1\\
\phi_2\\
     \end{array}
\right)
=
\left(
    \begin{array}{ccc}
 \frac{C_1}{C_D} & -\frac{C_2}{C_D}\\
-\frac{C_1}{C_S} & -\frac{C_2}{C_D}\\
     \end{array}
\right)
\left(
    \begin{array}{c}
\phi_1\\
\phi_2\\
     \end{array}
\right)
\label{Qandphi}
\end{eqnarray}
where $Q_\ell~ (\ell=D,S,G)$ are excess charges on the capacitors in Fig. 1 and
$\phi_\ell$ corresponding phases which are related to the potential difference $v_\ell(t)$
via the relation of ${\dot \phi}_\ell(t) =ev_\ell(t)/\hbar$.
Here, the capacitance $C_1$ and $C_2$ are defined by
$C_1=C_D C_S/(C_D+C_S)$ and $C_2=C_D^2 C_G/C_{\Sigma}(C_D+C_S)$.
Following Caldeira and Leggett\cite{caldeira}, one can express the circuit of Fig. 2
in terms of many  coupled harmonic oscillators each of which is quantized
under the commutation relation of $[\phi_j,Q_j]=ie$.
So, considering classical fields of voltage sources applied to the lumped circuit,
the electrical environment is described by a set of driven and coupled harmonic oscillators.

The tunneling part of the Hamiltonian may be given as,\cite{schoeller}
\begin{eqnarray}
{\cal H}'_{T}(t)\! &=&\!\sum_{\ell=S,D}\sum_{n k,k'}  \Big [
T_{kk'}^{n\ell} a_{nkI} a_{nk' \ell}^\dagger \sigma_+ e^{-i\phi_\ell}
\!+\!{\rm H.c.} \Big ]
\end{eqnarray}
where  $T_{kk'}^{n\ell}$ denotes an element of tunneling matrix between a state 
$\mid\! n k\rangle$ in the lead $\ell$
and a single particle state $\mid\! n k'\rangle$ in the island,
and usually approximated as 
$T_{kk'}^{n\ell} \simeq T^\ell$ independently of energy levels.
Here, the operators of $\sigma_+$ and $e^{i\phi_\ell}$ are inserted for
the increase of excess charges in the island and
the lead $\ell$, respectively.
From the commutation relation 
the operator $e^{i\phi_\ell}$ can be shown to increase excess charges
by the elementary charge $e$ in the lead $\ell$ for every tunneling event.

\subsection{Current and self-consistent calculations}
The current $I_\ell(t)$ flowing in each lead of Fig. \ref{setFig} is defined to be positive
if it flows into the island, so that some of charges carrying the current are used to increase charges
on the capacitor connected to the lead  while the others tunnel into the metallic island.
The former represents the displacement current $I^d_\ell(t)$ which is equal to
a time-derivative of the averaged charge as,
\begin{eqnarray}
I^d_\ell(t) = \frac{d}{dt}\langle Q_\ell(t) \rangle_0
\end{eqnarray}
where $Q_\ell(t)$ is the Heisenberg representation of the charge operator $Q_\ell$
and $\langle\cdots\rangle_0$ stands for the ensemble average.
Whereas, the latter is the tunneling current $I^t_\ell(t)$ which is equal to 
a time-derivative of the averaged particle number as,
\begin{eqnarray}
I^t_\ell(t) = e\frac{d}{dt}\langle {\cal N}_\ell(t) \rangle_0
\label{curr02}
\end{eqnarray}
where ${\cal N}_\ell(t)$ is the Heisenberg representation of the number operator
${\cal N}_\ell=\sum_{nk} a^\dagger_{nk\ell}a_{nk\ell}$ at the lead $\ell$ (we assume particles as
electrons).
In other words, the current $I_\ell(t)$ can be regarded as the sum of the displacement current
and the particle current which are contributed from 
the macroscopic and microscopic system, respectively,
\begin{eqnarray}
I_\ell(t) = I^d_\ell(t)+I^t_\ell(t) ~~~~(\ell=D,S,G),
\label{curr0}
\end{eqnarray}
with $I_G^t=0$. These currents  automatically satisfy the current conservation relation
of $I_D(t)+I_S(t)+I_G(t)=0$ despite of time-dependent perturbations, as emphasized
in Ref. \cite{buttiker}. This is another consequence of the charge conservation that
a Gaussian surface enclosing three capacitors defining the island always contains zero total charges
as inferred from Eq. (\ref{Q}). Then, from the continuity equation for charges,
the sum of the currents should be zero, alternatively, $d\langle Q(t)\rangle dt=I_D^t(t)+I_S^t(t)$
implying the fact that the increase of charges in the island is
enabled only by tunneling processes.
From the Heisenberg equation of motion for $Q(t)$,
one can show that this is the case for our system.

As for the displacement currents, it is possible to obtain further analytic forms
because the macroscopic system of ${\cal H}_{RLC}$ consists of linear elements.
By viewing the tunneling currents as another external sources as well as voltages of $V_D(t)$ and $V_G(t)$,
from the Heisenberg equations of motion for $Q_\ell(t)$ it is straightforward to show that,
\begin{eqnarray}
\left(
\begin{array}{cc}
 \frac{1}{i\omega C_D}+Z_D & -\frac{1}{i\omega C_S} \\
-\frac{1}{i\omega C_G}-Z_G & -\frac{C_S+C_G}{i\omega C_S C_G}-Z_G \\
\end{array}
\right)
\left(
\begin{array}{c}
 {\tilde I}^d_D(\omega)\\
 {\tilde I}^d_S(\omega)\\
\end{array}
\right)\nonumber\\
=
\left(
\begin{array}{c}
 {\tilde V}_D(\omega)-{\tilde V}_D^t(\omega) \\
 {\tilde V}_G(\omega)-{\tilde V}_G^t(\omega) \\
\end{array}
\right)
\label{eq:dcurrent}
\end{eqnarray}
where each current is expressed in its Fourier component defined by
$I(t) = \int e^{i \omega t} {\tilde I}(\omega)d \omega$.
Here, the voltages $V_D^t$ and $V_G^t$ are given by
\begin{eqnarray}
\left(
\begin{array}{c}
 {\tilde V}_D^t(\omega)\\
 {\tilde V}_G^t(\omega)\\
\end{array}
\right)=
\left(
\begin{array}{cc}
 Z_D & 0 \\
-\frac{1}{i\omega C_G}-Z_G &-\frac{1}{i\omega C_G}-Z_G\\ 
\end{array}
\right)
\left(
\begin{array}{c}
 {\tilde I}^t_D(\omega)\\
 {\tilde I}^t_S(\omega)\\
\end{array}
\right)
\label{Vteff}
\end{eqnarray}
describing effective voltage lowering by tunneling currents.

With the above expressions for the displacement currents,
the problem is now reduced to obtaining the tunneling currents of Eq. (\ref{curr02}).
For this, it is convenient to transform the system in such a way that 
the electrical environment of ${\cal H}_{RLC}$ leaves in a stationary condition.
Then, under such a situation, harmonic oscillators in ${\cal H}_{RLC}$ are expected to
vibrate about their stationary positions. This in turn is
helpful to assume them in equilibrium independently of tunneling events
even though noises from tunneling may modify their fluctuation slightly.
According to Ref. \cite{oh}, it is possible to find the unitary transformation which rotates
the system by voltages of $\delta V_D$ and $\delta V_G$.
By these voltages we mean that the system is rotated to have 
the lowered voltages in the macroscopic system by the amounts while
correspond phases in the tunneling Hamiltonian appear additionally.
Namely, the rotated Hamiltonian becomes,
\begin{eqnarray}
{\cal H}_{R}\!&=&\!{\cal H}_0+{\cal H}_T, \nonumber\\
{\cal H}_{0}\!&=&\!{\cal H}_{lead}+{\cal H}_I+{\cal H}_{RLC}(V_D-\delta V_D,V_G-\delta V_G),\nonumber\\
{\cal H}_T\!&=&\!\sum_{\ell=D,S}
\sum_{n k,k'}  \Big [
T^{\ell} a_{nkI} a_{nk' \ell}^\dagger \sigma_+ e^{-i\phi_\ell-i p_\ell(t)}
\!+\!{\rm H.c.} \Big ].
\end{eqnarray}
Here, additional terms of $p_\ell(t) (\ell=D,S)$ in the tunneling Hamiltonian actually describe
the external phase difference
forced by the voltages $\delta V_D$ and $\delta V_G$ in the absence of tunneling.
In other words, it is related to the corresponding potential difference $v_\ell^b(t)$
across the tunneling barrier from the island to the lead $\ell$
via $p_\ell(t)= e/\hbar \int^t_0 d\tau v_\ell^b(\tau)$.
The potential difference $v_\ell^b(t)$ is given by,
\begin{eqnarray}
\left(
\begin{array}{cc}
 \frac{C_S+C_G}{C_\Sigma} &-\frac{C_D}{C_\Sigma} \\
-\frac{C_G}{C_\Sigma}     & -\frac{C_G}{C_\Sigma}\\
\end{array}
\right)
{\bf Z}^{-1}
\left(
\begin{array}{c}
\tilde{v}^b_D(\omega)\\
\tilde{v}^b_S(\omega)\\
\end{array}
\right)
=
\left(
\begin{array}{c}
\frac{\delta{\tilde V}_D}{Z_D} \\
\frac{\delta{\tilde V}_G}{Z_G} \\
\end{array}
\right)
\label{vrb}
\end{eqnarray}
where the impedance matrix is defined as,
\begin{eqnarray}
{\bf Z}(\omega) \!=\! 
{\bf m} \left(
\begin{array}{cc}
i \omega C_1\!+\!Z_D^{-1}\!+\!Z_G^{\prime-1} & Z_G^{\prime-1} \\
Z_G^{\prime-1} & i \omega C_2\!+\!Z_G^{\prime-1} \\
     \end{array}
\right)^{-1}\!{\bf m}^T
\label{ReZ}
\end{eqnarray}
with $Z^\prime_G = Z_G (1+C_S/C_D)^2$.
Since the number operator ${\cal N}_\ell$ is found to be invariant under the rotation,
the transformation by $(\delta V_D,\delta V_G)=(V_D-V_D^t,V_G-V_G^t)$ may give rise to 
the simplest situation in calculating the tunneling currents.
In this case, the electrical environment is in stationary conditions as implied from Eq. (\ref{Vteff}),
so that the first moment of its dynamics does not influence the tunneling currents,
but the second moment, at least, starts to work.

Instead of this benefit, the rotated Hamiltonian now depends on the tunneling currents.
This implies that observables from the Hamiltonian also depends on the tunneling currents
unless both $V_D^t$ and $V_G^t$ in Eq. (\ref{vrb}) are zero.
Especially, in cases of the tunneling currents this requires self-consistent calculations to obtain it.
Detailed forms in the rotated frame become, from the Heisenberg equation of motion for  ${\cal N}_\ell$,
\begin{eqnarray}
\begin{array}{l}
I^t_\ell(t) = \langle  {\cal I}_\ell (t)\rangle_0, \\
{\cal I}_\ell(t) = {\cal  U}^\dagger(t,-\infty) {\cal J}_\ell(t) {\cal U}(t,-\infty), \\
{\cal J}_\ell(t)= \sum_{nkk'} \Big \{ \frac{e}{i\hbar}
T^\ell a_{nkI} a_{nk'\ell}^\dagger \sigma_+ e^{-i\phi_\ell-i p_\ell(t)}
\!+\!{\rm H.c.} \Big \}
\label{It}
\end{array}
\end{eqnarray}
showing self-consistent behavior.
Here,
the time-evolution operator ${\cal U}(t,t_0)$ is defined  in the rotated frame as,
\begin{eqnarray}
{\cal U}(t,t_0) = {\cal T} \exp\Big (\frac{1}{i\hbar}\int_{t_0}^t
d\tau {\cal H}_R(\tau) \Big ),
\label{evolution}
\end{eqnarray}
where ${\cal T}$ is the time-order operator.
Once the tunneling currents are calculated self-consistently, other observables
depend on them explicitly. For example, current noises are calculated as,
\begin{eqnarray}
S_{\ell \ell'}(t,t_0) =\langle \{ \delta {\cal I}_\ell(t),
\delta {\cal I}_{\ell'}(t_0)\} \rangle_0 
\label{noise}
\end{eqnarray}
which is defined by 
the auto-correlation function of the current fluctuation operator,
$\delta {\cal I}_\ell(t) = {\cal I}_\ell(t) -\langle {\cal I}_\ell(t)\rangle_0$.

\subsection{Equilibrium properties of reservoirs}

As shown in the unperturbed Hamiltonian, our system consists of  three
fermionic $({\cal H}_\ell,\ell=D,S,I)$ and one bosonic $({\cal H}_{RLC})$ systems.
We assume these systems as in equilibrium independently of tunneling because,
due to large degrees of freedom,  effects of tunneling on their fluctuation
are expected to be negligible.
Then, for the non-interacting fermionic systems, their dynamics are characterized by
single-particle Green's functions.
For a $\mid n\rangle$ state with its energy $\epsilon_n$ 
their explicit forms are as follows;
\begin{eqnarray}
g^K_{n\ell}(t,t')\!&\equiv&\! \frac{1}{i\hbar}
\langle [a_{n\ell}(t),a_{n\ell}^\dagger(t')]\rangle_0\nonumber\\
\!&=&\!\frac{1}{i\hbar} {\rm tanh}\frac{\beta \epsilon_{n\ell}}{2} e^{\epsilon_{n\ell}(t-t')/i\hbar}\nonumber\\
g^R_n(t,t') &\!\equiv\!& \frac{1}{i\hbar}\theta(t\!-\!t')
\langle\{a_n(t),a_n^\dagger(t')\}\rangle_0 \nonumber\\
\!&=&\!\frac{1}{i\hbar}\theta(t\!-\!t') e^{\epsilon_{n\ell}(t-t')/i\hbar}\nonumber\\
g^A_n(t,t') &\!\equiv\!& g^{R*}_n(t',t)
\label{g}
\end{eqnarray}
where $\beta=1/k_BT$ is inverse thermal energy and
$\epsilon_{n\ell}=\epsilon_n-\mu_\ell^0$ with
$\mu_\ell^0$ an equilibrium chemical potential at a lead $\ell$.\cite{chou,negele,utsumi}

As for the bosonic system of ${\cal H}_{RLC}(V_D^t,V_G^t)$,
it is now under a stationary condition;
$\langle \phi_i(t)\rangle_0 = \langle Q_i(t)\rangle_0 = 0$.
Then, its dynamical behavior is characterized by 
time-correlation functions between variables such as $\langle \phi_\ell(0) \phi_{\ell'}(t) \rangle_0$.
In thermal equilibrium, the time-correlation functions are easily evaluated
by exploiting the fluctuation-dissipation theorem,
and results are,\cite{ingold,oh}
\begin{eqnarray}
\langle\phi_\ell(0)\phi_{\ell'}(t)\rangle_0 =
2\int_{-\infty}^{\infty}\! \frac{d\omega}{\omega}
\frac{\Re  Z_{\ell\ell'}(\omega) }{R_K}
\frac{e^{i\omega t}}{1-e^{-\hbar\omega \beta}}
\label{fdiss}
\end{eqnarray}
where $Z_{\ell\ell'}(\omega)$ is related to each component of 
the impedance matrix ${\bf Z}(\omega)$ in such a way of $Z_{DD}={\bf Z}_{11}$,
$Z_{DS}={\bf Z}_{12}$, etc.

\section{ Statistical averages of operators }

\subsection{ Generating functional }
To evaluate the ensemble averages for the tunneling currents and noise
of Eqs. (\ref{It}) and (\ref{noise}),
we use the Schwinger-Keldysh approach combined with the generating functional
technique.\cite{utsumi,chou,babi}
We are interested in calculating the expectation value of ${\cal  O}$ defined by,
\begin{eqnarray}
\langle {\cal O}(t) \rangle_0 =
{\rm Tr}\{ \rho_0~ {\cal  U}^\dagger(t,-\infty)~ {\cal O}~
{\cal U}(t,-\infty) \}
\label{Obar}
\end{eqnarray}
where $\rho_0$ is the grand canonical density operator describing the system
in equilibrium at $t=-\infty$ as,
\begin{eqnarray}
\rho_0 = \exp\{-\beta( {\cal H}_0 -
\sum_{\ell=D,S,I} \mu^0_\ell{\cal N}_\ell)\} /Z_0,
\end{eqnarray}
and $Z_0$ is the equilibrium partition function.
In order to evaluate the expectation value of Eq. (\ref{Obar}),
we introduce a generating functional $W = -i\hbar~ {\rm ln} Z$
as an extension of the Gibbs free energy.
Here, $Z$ is  the generalized partition function defined as,
\begin{eqnarray}
Z = {\rm Tr}\{ \rho_0~ {\cal  U}_-^\dagger(-\infty,\infty)
{\cal U}_+(\infty,-\infty) \}
\end{eqnarray}
where, by different subscripts in the forward ${\cal U}_+(\infty,-\infty)$ and
backward ${\cal U}^\dagger_-(-\infty,\infty)$
evolution operators, we mean different external
fields applied along each time-branch, respectively.
Such different fields are usually coupled to the conjugate variable of ${\cal O}$
in the Hamiltonian and, at the final stage of calculations, are set to be identical.
A more compact form of the partition function is
obtained if we view the inverse temperature
as an imaginary time like,
\begin{eqnarray}
\rho_0 &=& \exp\Big\{\frac{1}{i\hbar}\int_{-\infty}^{-\infty-i\hbar\beta}
d\tau( {\cal H}_0 -\sum_{\ell=D,S,I} \mu_\ell{\cal N}_\ell) \Big\}
\nonumber\\
&\equiv&
{\cal U}_\tau(-\infty-i\hbar\beta,-\infty).
\end{eqnarray}
Here, in accordance with Eq. (\ref{evolution}), we define this density operator
with the evolution operator ${\cal U}_\tau$.
Then, the partition function becomes,
\begin{eqnarray}
Z &=& {\rm Tr}\{ {\cal U}_\tau(-\infty\!-\!i\hbar\beta,-\infty) 
{\cal  U}_-^\dagger(-\infty,\infty)   {\cal U}_+(\infty,-\infty) \}
\nonumber\\
&\equiv& {\rm Tr} \{ {\cal U}_C \},
\end{eqnarray}
and can be interpreted as describing
successive evolutions of states along $C_+$-, $C_-$-, and
$C_\tau$-time branches as shown in Fig. 3.
We designate these time branches  simply
a closed-time path $C$ bearing in mind that,
along each time branch,
different Hamiltonians govern the evolution of states;
that is, ${\cal H}_0+{\cal H}_T$ along the $C_\pm$-time branches
with different fields and ${\cal H}_0-\sum_\ell \mu_\ell {\cal N}_\ell$
along the $C_\tau$-time branch.

Once the generating functional $W$ is given, the ensemble average of
Eq. (\ref{Obar}) is obtained through functional derivatives of $W$
with respect to external fields.
In cases of the tunneling currents and their noises, the phases $p_\ell(t)$ are assumed to
be different along each time branch such as,
\begin{eqnarray}
p_\ell^+(t) \!&=&\! p_\ell(t)\!+\!\frac{\Delta p_\ell(t)}{2}
~~~~{\rm for~forward}\nonumber\\
p_\ell^-(t) \!&=&\! p_\ell(t)\!-\!\frac{\Delta p_\ell(t)}{2}
~~~~{\rm for~backward}
\label{pfic}
\end{eqnarray}
where $\Delta p_\ell(t)$ is fictitious and will be zero at the final stage.
Then, the functional derivatives of the evolution operators with respect to
the fictitious field give,
\begin{eqnarray}
\left. \frac{\delta {\cal U}_+(\infty,-\infty) }{\delta \Delta p_\ell(t) }\right
|_{\Delta p_\ell\!=\!0}
\!&=&\!\frac{i}{2e} {\cal U}(\infty,t) {\cal J}_\ell(t) {\cal U}(t,-\infty)
\nonumber\\
\left. \frac{\delta {\cal U}_-^\dagger(\infty,-\infty) }{\delta \Delta p_\ell(t) }\right
|_{\Delta p_\ell \!=\!0}
\!&=&\!\frac{i}{2e} {\cal U}(-\infty,t) {\cal J}_\ell(t) {\cal U}(t,\infty)
\end{eqnarray}
and, using these relations,
it is straightforward to show the tunneling current of Eq. (\ref{It}) to be,
\begin{eqnarray}
I_\ell^t(t) = \frac{e}{\hbar} \left. \frac{\delta W}{\delta \Delta p_{\ell}(t)}
\right |_{\Delta p_{\ell}=0}.
\label{WIt}
\end{eqnarray}
In a similar way, the current noises are obtained by
the second derivatives,
\begin{eqnarray}
S_{\ell \ell'}(t,t_0)\!=\! -\frac{e^2}{i\hbar}
\Big \{
 \frac{\delta^2 W}{\delta p_\ell^-(t) \delta p^+_{\ell'}(t_0)}
\!+\!\frac{\delta^2 W}{\delta p_\ell^+(t) \delta p^-_{\ell'}(t_0)}
\Big \}\nonumber\\
\!=\!  \frac{2e^2}{i\hbar}
\Big  \{
 \frac{\delta^2 W}{\delta \Delta p_\ell(t) \delta \Delta p_{\ell'}(t_0)}
-\frac{1}{4}
 \frac{\delta^2 W}{\delta p_\ell(t) \delta p_{\ell'}(t_0)}
\Big \}
\label{WSaa}
\end{eqnarray}
accompanied with $\Delta p_\ell(t)=\Delta p_\ell(t_0)=0$ finally.
Due to the normalization of the partition function $Z(\Delta p_\ell=0)=1$,
the second term in the last line of the above equation is equal to  zero.
Nevertheless, we keep this term to circumvent the uncertainty related to the order
of operators.\cite{utsumi}
\begin{figure}
\includegraphics[width=0.4\textwidth]{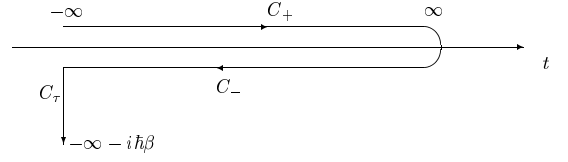}
\caption{\label{fig3}
Time contour to evaluate the partition function}
\end{figure}

In order to get the averaged charge in the island and its fluctuation,
we add a fictitious field to the excitation energy in such a way of
$\Delta_0\pm \Delta h(t)/2$.
Then, using similar procedure for the tunneling currents,
one gets the ensemble average of a charge operator $Q$ as,
\begin{eqnarray}
\langle Q (t)\rangle_0= -\left. \frac{e \delta W}{\delta \Delta h(t)}
\right |_{\Delta h=0}
\label{WQ}
\end{eqnarray}
while its fluctuation is given by,
\begin{eqnarray}
S_{Q}(t,t_0) = \langle \{\delta Q(t), \delta Q(t_0)\} \rangle_0
~~~~~~~~~~~~~~~~~~~~~~~~~~~~~~ \nonumber\\
= 2 e^2 \left (
\frac{1}{4} \frac{i\hbar \delta^2 W}{ \delta \Delta_0(t) \delta \Delta_0(t_0) }
-\frac{i\hbar \delta^2 W}{ \delta \Delta h(t) \delta \Delta h(t_0) }
\right )_{\Delta h=0}.
\label{WSqq}
\end{eqnarray}
Here, we define the charge fluctuation operators as
$\delta Q(t) = Q(t) -\langle  Q(t) \rangle_0$.

\subsection{ Evaluation of Generating functional }

We evaluate the generating functional $W$ with the coherent-state functional
integral method defined on the closed time-path of Fig. 3.
As an usual path integral,
the evolution operator ${\cal U}_C(t)$ is divided into a number of 
infinitesimal steps on the time-path $C$, and then
a resolution of the identity is inserted at every time step.
In the coherent state functional integral differently from usual ones,
the identity operator is given 
in terms of eigenfunctions of annihilation operators instead of
coordinate- and phase-state basis functions.
As a consequence,
the evaluation of the partition function $Z$ is
reduced to the path integral in coherent-state variables over the exponential of the action
along the time contour $C$.\cite{negele}
In our case, the result is summarized as
\begin{eqnarray}
Z = Z_0
\langle \langle \langle
e^{ -\frac{1}{i\hbar}S_{int} } \rangle_{{\cal H}_{lead}}
\rangle_{{\cal  H}_I} \rangle_{{\cal  H}_{RLC}} \label{Z}
\label{Zformal},
\end{eqnarray}
in other words, the exponential of the action $S_{int}$
is averaged over all reservoirs.
The action $S_{int}$ describes the interaction among reservoirs,
which is given by $S_{int}= \oint_C d\tau H_T(\tau)$
on the closed time-path $C$.
Here, $H_T(t)$ is a counterpart of the tunneling Hamiltonian
for each time branches, and is obtained by
replacing all operators 
with their coherent-state  variables (complex or Grassman numbers)
while  $H_T(t)= 0$ in the time branch $C_\tau$.
As for spins, in order to utilize the coherent-state representation,
we map the effective spin-1/2 operators onto
two fermion operators $c$ and $d$ (drone-fermion representation),
i.e., $\sigma_+= c^\dagger(d^\dagger+d)$ and $\sigma_z=2c^\dagger c-1$.\cite{utsumi}

By the bracket notation, we mean an average weighted with the action of
a certain reservoir.
The detailed form is defined, for instance over
${\cal H}_{lead}$, as
\begin{eqnarray}
\langle O \rangle_{{\cal H}_{lead}} = \frac{1}{Z_0^{lead}}
\int {\cal D}[a_{rk\ell}(t),a^*_{rk\ell}(t)]
e^{ -\frac{1}{i\hbar} S_{lead} }~ O
\end{eqnarray}
where $Z_0^{lead}$ is a normalization factor
implying $\langle 1 \rangle_{{\cal H}_{lead}} = 1$ and equal
to the equilibrium partition function of ${\cal H}_{lead}$.
Here, $\{a_{nk\ell}(t),a_{nk\ell}^*(t)\}$ are Grassmann variables
associated with their fermionic operators and they
satisfy anti-periodic boundary conditions; $a_{nk\ell}(-\infty\in C_+)
=-a_{nk\ell}(-\infty-i\hbar\beta \in C_\tau)$.
The unperturbed action of the leads $S_{lead}$ is given as,
\begin{eqnarray}
S_{lead} = i\hbar \oint_C d\tau \Big \{
\sum_{nk\ell}
a^*_{nk\ell}(\tau) \frac{\partial}{\partial\tau} a_{kn\ell}(\tau)
      \!-\!H_{lead}(\tau)  \Big \}
\end{eqnarray}
using the trajectory notation, in which
the function $H_{lead}(t)$ represents ${\cal H}_{lead}$
and ${\cal H}_{lead}-\sum_\ell \mu_\ell {\cal N}_\ell$
in the time branches $C_\pm$ and $C_\tau$ respectively.

Since the unperturbed actions are quadratic, 
the thermal average over all reservoirs of Eq. (\ref{Zformal}) is reduced to
 Gaussian times polynomial integrals if one expands $e^{-S_{int}/i\hbar}$ into a power series.
Then, using a standard procedure of a Gaussian integral, each term in the series can be
evaluated analytically.
Firstly the result over the reservoirs $(\ell=D,S,I)$ is summarized by
the appearance of a particle-hole Green's function $b_\ell(t,t')$ in the action $S_{int}$.
The Green's function $b_\ell(t,t')$ has a form of,
\begin{eqnarray}
b_\ell(t,t_0) &=& -i\hbar N_{ch} \sum_{k}\mid T^{\ell}\mid^2
g_{kI}(t,t_0) g_{k\ell}(t_0,t)\nonumber\\
&& e^{i p_\ell(t)-i p_\ell(t_0)}, ~~~~~\ell=D,S
\end{eqnarray}
Here, the free-particle Green's functions $g_{k\ell}(t,t_0)$
represents the inverse function of their free actions 
and their physical representations are equal to Eq. (\ref{g})
which can be obtained by the Keldysh rotation.
Then, the function $b_\ell(t,t')$
represents just a single particle or hole creation in the island by tunneling
through a barrier $\ell$.
Actually, in obtaining $b_\ell(t,t')$ we take into account only
sequential processes of single particle or hole creation; one can see it
if expanding the exponential of $S_{int}$ into a series.
This is a good approximation for a large number of transverse channels $N_{ch}$, 
so called, the wide junction limit because the sequential particle or hole creation is
dominated to the simultaneous creation of both, at least, by $N_{ch}$.\cite{schoeller}

These sequential processes are correlated by further evaluation of the partition function
over the $c-$ and $d-$ fields.
The result reads,
\begin{eqnarray}
Z = Z_0 \left\langle \exp \Big[-\sum_{n=1} \frac{1}{n} {\rm Tr} \{ 
(g_c\frac{\delta}{\delta \eta} B \frac{\delta}{\delta \eta})^n \} \Big]
\right \rangle_{{\cal H}_{RLC}}\nonumber\\
\left. \exp\{ -i\hbar \int_C dt dt_0~ \eta(t) g_d(t,t_0)\eta(t_0) \}
\right |_{\eta =0}
\label{Z1}
\end{eqnarray}
where
the simplified notation of 
${\rm Tr}\{ g_c\frac{\delta}{\delta\eta} B \frac{\delta}{\delta\eta}\}$ 
stands for
\begin{eqnarray}
{\rm Tr}\{ g_c\frac{\delta}{\delta\eta} B
\frac{\delta}{\delta\eta}\} \!=\!
\int_C\! dt\! dt_0
g_c(t,t_0)\frac{\delta}{\delta\eta(t_0)} B(t_0,t) \frac{\delta}{\delta\eta(t)}
\end{eqnarray}
and
$B$ is the particle-hole Green's function
combined with effects of the electrical environment by
\begin{eqnarray}
B(t,t_0)=\sum_\ell e^{-i\phi_\ell(t)}
b_\ell(t,t_0) e^{i\phi_\ell(t_0) }.
\label{B}
\end{eqnarray}
Here, $g_c$ and $g_d$ are equilibrium Green's functions for the $c-$ and $d-$ fields 
with their eigenenergies $\Delta_0$ and zero, respectively.
The Grassmann field $\eta(t)$ is introduced as a linear source
to the $d$-field, which gives  an additional term 
of $-i\hbar \int_C dt [ \eta(t) d(t)-d^*(t) \eta(t) ]$ into the unperturbed action.

Finally, we evaluate the partition function over the electrical environment.
To do this, we transform bi-linearly coupled harmonic oscillators in 
${\cal H}_{RLC}$ into independent ones by an unitary transformation.
Then, since each simple harmonic oscillator is assumed to be in equilibrium,
one can exploit the Wick's theorem to show that its thermal average can be expressed
in terms of, at most, the second order phase correlation.
To each term of the series in Eq. (\ref{Z1}),
the application of the theorem leads to the relation of,
\begin{eqnarray}
\langle e^{i \phi_{\ell_1}(1)}
 e^{-i \phi_{\ell_2}(2)} \ldots e^{-i \phi_{\ell_n}(n)}
\rangle_{{\cal H}_{RLC}}
=
\nonumber\\
e^{ K_{\ell_1\ell_2\ldots\ell_n}(1,2,\ldots,n) }
\end{eqnarray}
where $K_{\ell_1\ell_2\ldots}(1,2,\ldots,n)$ is defined by
\begin{eqnarray}
K_{\ell_1\ell_2\ldots\ell_n}(1,2,\ldots,n) =~~~~~~~~~~~~~~~~~~~~~~~~~ \nonumber\\ 
\frac{1}{2}
\langle [ i\phi_{\ell_1}(1)\!-\!i\phi_{\ell_2}(2)\!+\ldots-i\phi_{\ell_n}(n) ]^2
\rangle_{{\cal H}_{RLC}}.
\end{eqnarray}
According to this, $K_{\ell_1\ell_2\ldots}(1,2,\ldots)$ is the sum of
the second-order correlation functions among all time-arguments, so that
all sequential processes of microscopic variables in each term of Eq. (\ref{Z1}) are
correlated additionally.
Actually, the second-order correlation function of
$\langle \phi_\ell(1)\phi_{\ell'}(2)\rangle_{{\cal H}_{RLC}}$
is related to the phase fluctuations of Eq. (\ref{fdiss}).
One of simple ways for this is to display the fluctuation-dissipation theorem
in terms of $K_{\ell_1\ell_2}(1,2)$ and then, by comparing it with Eq. (\ref{fdiss}), one obtains;
\begin{eqnarray}
\langle \phi_\ell(t)\phi_{\ell'}(t_0)\rangle_{{\cal H}_{RLC}}^{-+}
&=& \langle \phi_\ell(t)\phi_{\ell'}(t_0) \rangle_0  \nonumber\\
\langle \phi_\ell(t)\phi_{\ell'}(t_0)\rangle_{{\cal H}_{RLC}}^{+-}
&=& \langle \phi_\ell(t)\phi_{\ell'}(t_0)\rangle_0^* \nonumber\\
\langle \phi_\ell(t)\phi_{\ell'}(t_0)\rangle_{{\cal H}_{RLC}}^{++}
&=& \theta(t-t_0) \langle \phi_\ell(t)\phi_{\ell'}(t_0) \rangle_0 
\nonumber\\&+&\theta(t_0-t)
\langle \phi_\ell(t)\phi_{\ell'}(t_0) \rangle_0^* \nonumber\\
\langle \phi_\ell(t)\phi_{\ell'}(t_0)\rangle_{{\cal H}_{RLC}}^{--}
&=& \theta(t-t_0)
\langle \phi_\ell(t)\phi_{\ell'}(t_0) \rangle_0^*
\nonumber\\&+&\theta(t_0-t)
\langle \phi_\ell(t)\phi_{\ell'}(t_0) \rangle_0
\end{eqnarray}
where superscripts of $\pm\!\mp$ denote each section of the Keldysh space
in which both time arguments  $t$ and $t_0$ belong.

The generating functional $W$ is now calculated by summing
all-connected diagrams in $Z$.
To prevent the divergence of the average charge at $\Delta_0=0$, we perform
diagrammatic sum to infinite order, however
approximated forms in higher-order diagrams
cannot be inevitable for a simple form of $W$.
Expanding the partition function of Eq. (\ref{Z1}) and arranging diagrams,
we write the approximated generating functional as,
\begin{eqnarray}
W & \simeq & -i\hbar \left( {\rm Tr}[{\rm ln}g_c^{-1}]- {\rm Tr}[ g_c \Sigma^{(1)} ]
-\frac{1}{2} {\rm Tr}[ g_c \Sigma^{(2)} ] \right. \nonumber\\
&&\left. -\frac{1}{3} {\rm Tr}[ g_c \Sigma^{(3)} ] 
+\cdots \right )
\label{Wa}
\end{eqnarray}
where  we omit trivial non-interacting terms.
In the first-order term, a single diagram contributes the generating functional
and its self-energy has a form of,
\begin{eqnarray}
\Sigma^{(1)}(t,t_0)&=& \sum_{\ell}
e^{K_{\ell\ell}(t,t_0)} \Sigma^f_{\ell}(t,t_0)
\end{eqnarray}
with a free-environment part of,
\begin{eqnarray}
\Sigma^f_{\ell}(t,t_0) = -2i\hbar g_d(t_0,t) b_\ell(t,t_0).
\end{eqnarray}
However, for each higher-order term from the second, rich connected diagrams are found.
We approximate the generating function by taking into account only a single diagram in
each $n$\_order with its self energies of,
\begin{eqnarray}
\Sigma^{(n)}(t,t_0)= \sum_{\ell_1..\ell_n} \int_C d1 d2.. dn
e^{K_{\ell_1\ell_1\ldots\ell_n\ell_n}(t,1,2,..,t_0)} \nonumber\\
\Sigma^f_{\ell_1}(t,1) g_c(1,2)
\Sigma^f_{\ell_2}(2,3) g_c(3,4)..\Sigma^f_{\ell_n}(n,t_0).
\end{eqnarray}
The idea for such preferred diagrams comes from the work of Utsmi et. al.\cite{utsumi}
Actually, the partition function of $Z$ is the same as that in their work
if it were not for the electrical environment and time-dependent perturbations.
So we compose the approximated generating functional to recover their results
in the absence of the electrical environment and time-dependent perturbations, say,
in the case of $K_{\ell\ldots}=0$ and $p_\ell(t) = 0$.

As a result, the generating functional can be written in a more compact form as,
\begin{eqnarray}
W &=& -i\hbar\left ( {\rm Tr}[{\rm ln}g_c^{-1}]-\sum_{n=1}^\infty
\frac{1}{n} {\rm Tr}[ (g_c \Sigma )^n] \right )
\nonumber\\
  &=& -i\hbar~ {\rm Tr}[{\rm ln} G^{-1}]
\label{W2}
\end{eqnarray}
where we introduce the full $c$-field Green's function $G$
obeying the Dyson equation;
\begin{eqnarray}
G^{-1}(t,t_0) = g_c^{-1}(t,t_0)-\Xi(t,t_0)
\label{G}
\end{eqnarray}
with a noninteracting Green's function $g_c(t,t_0)$ of the $c-$ field.
Here, the self-energy $\Xi(t,t_0)$ is defined by,
\begin{eqnarray}
&&\Xi(t,t_0) = \Sigma^{(1)}(t,t_0)\nonumber\\
&&+\frac{1}{2} \left[ \Sigma^{(2)}-
\Sigma^{(1)}g_c \Sigma^{(1)} \right ](t,t_0)\nonumber\\
&&+\frac{1}{3} \left [ \Sigma^{(3)}
-\frac{3}{2}\Sigma^{(2)} g_c\Sigma^{(1)}
+\frac{1}{2}\Sigma^{(1)} g_c\Sigma^{(1)}g_c\Sigma^{(1)} \right](t,t_0)
\nonumber\\
&&+\ldots\nonumber\\
&&\simeq \Sigma^{(1)}(t,t_0)
\label{Sigma}
\end{eqnarray}
which is obtained by comparing Eqs. (\ref{Wa}) and (\ref{W2}).
Since Eq. (\ref{G}) already sum up an infinite series, we calculate $\Xi(t,t_0)$
to the lowest order of $\Sigma^{(1)}(t,t_0)$.
This is equivalent to neglecting additional correlations caused by the electrical environment
in higher-order terms of the above equation.
In this case, the self-energy of $\Xi(t,t_0)$ is expressed explicitly
with the product of terms representing effects of time-dependent perturbations
and the electrical environment, respectively, as
\begin{eqnarray}
\Xi(t,t_0) &=& 
\Xi_D(t,t_0)+\Xi_S(t,t_0), \nonumber\\
\Xi_\ell(t,t_0) &=& e^{i p_\ell(t)} \Sigma_{\ell}(t,t_0) e^{-i p_\ell(t_0)},\nonumber\\
\Sigma_\ell(t,t_0) &=& 
e^{K_{\ell\ell}(t,t_0)} \Sigma^0_{\ell}(t,t_0)
\label{Sigma3}
\end{eqnarray}
and thus $\Sigma^0_{\ell}$ represents the free self-energy from the environment and time-dependent
perturbations,
\begin{eqnarray}
\Sigma^0_{\ell}(t,t_0) &=& -2 \hbar^2 N_{ch} g_d(t_0,t) \mid T^\ell\mid^2 \nonumber\\
&&\sum_k g_{kI}(t,t_0) g_{k\ell}(t_0,t).
\label{Sigma3a}
\end{eqnarray}

\section{ Expressions for currents and noises }

For the given generating functional and the self-energy of Eqs. (\ref{W2}) and (\ref{Sigma3}),
we now derive exact expressions for tunneling currents, averaged charges, and their noises.
Using the standard procedure, we perform 
functional derivatives as specified in Eqs. (\ref{WIt}), (\ref{WSaa}), (\ref{WQ}),
and  (\ref{WSqq}), and transform them into the physical representation.
The transformations are carried out by adopting the Keldysh rotator as, for instance of $G$,
\begin{eqnarray}
\left(
    \begin{array}{cc}
 G^{+\!+} & G^{+\!-} \\
 G^{-\!+} & G^{-\!-}
     \end{array}
\right)
\!=\! \frac{1}{2}
\left(
\begin{array}{cc}
 1 & 1 \\
 -\!1& 1 
\end{array}
\right)
\left(
    \begin{array}{cc}
 0 & G^A \\
 G^R & G^K 
     \end{array}
\right)
\left(
\begin{array}{cc}
 1 & -\!1 \\
 1 & 1 \\
\end{array}
\right),
\end{eqnarray}
where superscripts $A$, $R$, and $K$ denote its advanced, retarded, and Keldysh
components, respectively.

Then, with this rotator  the Dyson equation of Eq. (\ref{G}) is transformed as;
\begin{eqnarray}
G^R(t,t_0) = g_c^R(t,t_0)~~~~~~~~~~~~~~~~~~~~~~~~~~~~~~~~~~~~~~~~ \nonumber\\
+ \int_{-\infty}^\infty  d\tau_1 d\tau_2 g_c^R(t,\tau_1)\Xi^R(\tau_1,\tau_2)
G^R(\tau_2,t_0) \nonumber\\
G^K(t,t_0) \!=\! \int_{-\infty}^\infty\!  d\tau_1\! d\tau_2 G^R(t,\!\tau_1)
\Xi^K(\tau_1,\!\tau_2) G^A(\tau_2,\!t_0)\nonumber\\
G^A(t,t_0) \!=\! G^{R*}(t_0,t)~~~~~~~~~~~~~~~~~~~~~~~~~~~~~~~~~~~~~~
\label{G2}
\end{eqnarray}
where $g_c^R(t,t_0)$ is a retarded component of the free-particle Green's function
specified in Eq. (\ref{g}).  According to these relations, $G^{R,A}$ has the same
causality relation as that of $g_c^{R,A}$, and additionally satisfies
the sum rule of $\lim_{t\rightarrow t_0} i\hbar G^C(t,t_0)=1$ with $G^C \equiv G^R-G^A$.
For $G^K$, we show it in terms of $G^R$ and $G^A$ rather than its 
integral equation by noting that $({\bf 1}+G^R\Sigma^R)
g_c^K = G^R(g_c^R)^{-1} g_c^K$ vanishes.
On the other hand, the application of the rotator to the self-energy  of Eq. (\ref{Sigma3}) leads to
\begin{eqnarray}
\begin{array}{l}
\Sigma^{R}_\ell(t,t_0) = \Sigma^{0R}_{\ell}(t\!-\!t_0) \Re e^{K^{-+}_{\ell\ell}(t,t_0)} \\
~~~~~~~~~~~+i \Sigma^{0K}_{\ell}(t\!-\!t_0) \Im e^{K^{-+}_{\ell\ell}(t,t_0)} \theta(t\!-\!t_0),\\
\Sigma^{K}_\ell(t,t_0) = \Sigma^{0K}_{\ell}(t\!-\!t_0) \Re e^{K^{-+}_{\ell\ell}(t,t_0)}  \\
~~~~~~~~~~~+ i \Sigma^{0C}_{\ell}(t\!-\!t_0) \Im e^{K^{-+}_{\ell\ell}(t,t_0)},\\
\Sigma^{A}_\ell(t,t_0) = \Sigma^R_\ell(t_0,t)^*.
\end{array}
\end{eqnarray}
Here, we arrange each term to depend only on the correlation function of
$K^{-+}_{\ell\ell}(t,t_0)$. Thus, the self-energy of $\Sigma_\ell$ is
directly related to the phase fluctuation of Eq. (\ref{fdiss}) via
$K^{-+}_{\ell\ell}(t,t_0) = \langle \phi_\ell(0)[\phi_\ell(t_0-t)-\phi_\ell(0)] \rangle_0$.

A detailed form of the bare self-energy $\Sigma^0_{\ell}$ depends on
a cut-off function for the spectral density of the particle-hole propagator $b_\ell$.
In this work we use a Lorenzian cut-off function $\rho(E)=EE_0^2/(E^2+E_0^2)$
with a bandwidth of $E_0=E_C$ as in earlier works.\cite{schoeller,utsumi}
Then, by substituting free-particle Green's functions
and $\rho(E)$ into (\ref{Sigma3a}), and transforming into Fourier forms by
\begin{eqnarray}
\Sigma^0_\ell(t,t_0) = \frac{1}{2\pi }
\int d\omega e^{-i\omega(t-t_0) } \tilde\Sigma^0_\ell(\hbar\omega),
\end{eqnarray}
we find,
\begin{eqnarray}
\tilde\Sigma^{0R}_{\ell}(\epsilon) &=&  \frac{R_K}{4\pi^2 R_\ell}
\rho(\epsilon_\ell) \Big [ 2\Re~ \psi_0(1\!+\!i \frac{\epsilon_\ell\beta}{2\pi}
)\nonumber\\
&-&\psi_0(1\!+\!\frac{E_0\beta}{2\pi}) -\psi_0(\frac{E_0\beta}{2\pi})\!-\!i \pi
{\rm coth}\frac{\beta \epsilon_\ell}{2}  \Big ], \nonumber\\
\tilde\Sigma^{0K}_{\ell}(\epsilon) &=& i \frac{R_K}{2\pi R_\ell} \rho(\epsilon_\ell),\nonumber\\
\tilde\Sigma^{0A}_{\ell}(\epsilon) &=& \tilde\Sigma^{0R*}_{\ell}(\epsilon)
\label{Sigma0}
\end{eqnarray}
where $\epsilon_\ell=\epsilon-\mu_\ell^0$ and $\psi_0$ is a digamma function.
Here,
$R_\ell$ stands for $1/R_\ell = 4 \pi^2 N_{ch} \mid\! T^\ell \!\mid^2 D_I D_\ell/R_K$
with energy-independent density of states $D_I$ and $D_\ell$ at the metallic island and
the lead $\ell$, respectively.
This is  tunneling resistance of the barrier connected to the lead $\ell$,
so that the parallel resistance is given by $1/R_T = 1/R_D+1/R_S$
in $\alpha_0=R_K/(4 \pi^2 R_T)$.

For static conditions, an analytical solution of the Dyson equation in Eq. (\ref{G2}) is easily obtained by
transforming it to the Fourier space. However, for arbitrary time-dependent perturbation
since the time-translational invariance of the self-energy is broken,
the solution shows up in a series or it is necessary to solve the problem numerically.
One of approximated solutions may be derived by noting that $\Sigma^0_\ell(t,t_0)$ exhibits
rapid decaying behavior as a function of time interval $t-t_0$.
If this is the fastest time-variation in the problem, the integral equation can be approximated to the
first-order differential equation. Then, the solution becomes similar to that obtained
in the the wide-band limit.\cite{jauho}

With the physical representation of the Green's functions and the self-energies,
results of the functional derivatives are summarized as the followings.
The averaged charge in the island and the tunneling currents are given as,
\begin{eqnarray}
\langle Q (t) \rangle_0 \!&=&\! -e\frac{i\hbar}{2}  G^K(t,t),
\label{Qbar}\\
I_\ell^t(t)\! &=&\! -e \Re \Gamma_\ell^K(t,t)
\label{Itell}
\end{eqnarray}
while their fluctuations read 
\begin{eqnarray}
S_Q(t,t_0) &=& \frac{\hbar^2 e^2}{2} G^K(t,t_0) G^K(t_0,t)-(K\rightarrow C),
\label{Qnoise}\\
S_{\ell\ell'}(t,t_0) &=& e^2 \Re
\Big\{ \Xi^K_\ell(t,t_0)G^K(t_0,t)\delta_{\ell\ell'} \nonumber\\
\!&+&\!\Lambda_{\ell\ell'}^K(t,t_0) G^K(t_0,t)-\Gamma^K_\ell(t,t_0)\Gamma_{\ell'}^K(t_0,t) \Big\}
\nonumber\\ &-& (K\rightarrow C)
\label{Snoise}
\end{eqnarray}
where $(K\rightarrow C)$ means the change of a Keldysh component to a correlated one,
for example, $G^K\rightarrow G^C = G^R-G^A $.
Here, the functions of $\Gamma$ and $\Lambda$ are defined as,
\begin{eqnarray}
\Gamma_\ell^{K,C}(t,t_0) &=& \int_{-\infty}^\infty d\tau
\Big[ \Xi^{K,C}_\ell(t,\tau) G^A(\tau,t_0)\nonumber\\
&+& \Xi^R_\ell(t,\tau) G^{K,C}(\tau,t_0)\Big ],\nonumber\\
\Lambda_{\ell\ell'}^{K,C}(t,t_0) &=& \int_{-\infty}^\infty d\tau
\Big[ \Gamma^{K,C}_\ell(t,\tau) \Xi_{\ell'}^A(\tau,t_0) \nonumber\\
&+& \Gamma^R_\ell(t,\tau) \Xi_{\ell'}^{K,C}(\tau,t_0) \Big ],
\end{eqnarray}
with $\Xi^C_\ell = \Xi^R_\ell-\Xi^A_\ell$.
With these fluctuations, the noise spectrum at a frequency
$\omega$ is defined by,\cite{davies,hanke,korotkov}
\begin{eqnarray}
S_Q(\omega) &=& \langle\!\langle S_Q(t,t_0) \cos\omega (t-t_0) \rangle\!\rangle\nonumber\\
S_{\ell\ell'}(\omega) &=& \langle\!\langle S_{\ell\ell'}(t,t_0) \cos\omega (t-t_0) \rangle\!\rangle
\end{eqnarray}
where the double bracket denotes the integration of,
\begin{eqnarray}
\langle\!\langle ...  \rangle\!\rangle = \Re\left[ \lim_{T\rightarrow\infty}\frac{1}{T}\int_0^{T} dt_0
\int_{-\infty}^{\infty} dt ~... \right].
\label{daverage}
\end{eqnarray}
As discussed in the previous section the above results obey the charge conservation law even
under arbitrary time-dependent perturbations, i.e.
\begin{eqnarray}
\frac{ \partial \langle Q (t) \rangle_0 }{\partial t} &=& \sum_\ell I_\ell^t(t),
\\
\frac{ \partial^2 S_Q(t,t') }{\partial t \partial t'} &=&
\sum_{\ell\ell'} S_{\ell\ell'}(t,t').
\label{Scon}
\end{eqnarray}
This is also a direct consequence of the guage-invariant generating functional of Eq. (\ref{W2}).

The equations of Eq. (\ref{Qbar})-(\ref{Snoise}) are the main results of our work
together with the generalized self-energies of Eq. (\ref{Sigma3})
to arbitrary time-dependent perturbations and electrical environments.
Even though the above equations give the exact expressions within the given generating functional and
provide easier ways in numerical calculations,
their physical meanings are not well revealed.
So, in the subsequent section we present another forms
by considering various limits.

\subsection{ Wide-band limit }

In order to obtain more physically meaningful forms we start with 
defining a spectral function $A_\ell(\epsilon,t)$ as 
\begin{eqnarray}
A_\ell(\epsilon,t) = \int_{-\infty}^\infty d \tau e^{i\epsilon(\tau-t)/\hbar} 
e^{-i p_\ell(\tau) } G^A(\tau,t)
\end{eqnarray}
and writing ${\tilde \Sigma}^K_\ell(\epsilon)$ as,
\begin{eqnarray}
\tilde{\Sigma}^K_{\ell}(\epsilon)=\tilde{\Sigma}_{\ell}^C(\epsilon)
\{ 2f_\ell(\epsilon) -1 \} 
\label{FD}
\end{eqnarray}
where $f_\ell(\epsilon)$ is a Fermi-Dirac distribution function
broadened due to the presence of the dissipative environment, otherwise,
it is equal to 
 $f_\ell(\epsilon) = 1/(1+e^{\beta(\epsilon-\mu_\ell^0)})$.
Then, from Eq. (\ref{Qbar}) the average charge can be rewritten as,
\begin{eqnarray}
\langle Q (t) \rangle_0 &=& e\sum_\ell \int d \epsilon~
n_\ell(t,t;\epsilon) \{ f^+_\ell(\epsilon) -f_\ell(\epsilon) \}
\label{Q2}
\end{eqnarray}
where, by defining  a hole-distribution function,
$f^+_\ell(\epsilon)=1-f_\ell(\epsilon)$,  we emphasize the roles of hole and
electron contributions.
Here, $n_\ell(t,t;\epsilon)$ is a diagonal value of,
\begin{eqnarray}
n_\ell(t,t_0;\epsilon) \equiv 
\frac{i}{4 \pi } e^{-i \epsilon(t-t_0)/\hbar}  {\tilde\Sigma}_{\ell}^C(\epsilon)
A_\ell^*(\epsilon,t) A_\ell(\epsilon;t_0),
\end{eqnarray}
which can be interpreted as, owing to its unit,
the evolution of density of states at the lead $\ell$ available to occupying the island.

As for the tunneling currents, we additionally exploit the
relation of $(1+\Xi^R G^R) = (i\hbar\partial_t -\Delta_0)G^R$ and then
obtain the form of, from Eq. (\ref{Itell}),
\begin{eqnarray}
I_\ell^t(t) = \frac{1}{2eR_K} \Re \!\int_{-\infty}^\infty d\epsilon \Big[
~~~~~~~~~~~~~~~~~~~~~~~~~~~~~~~~~\nonumber\\
\left\{ T^F_\ell(t,t;\epsilon)\!-\!T^R_\ell(t,t;\epsilon)\right\}
\left\{f^+_\ell(\epsilon) \!-\! f_\ell(\epsilon)\right\} \nonumber\\
-T^F_{\bar\ell}(t,t;\epsilon)
\left\{ f^+_{\bar \ell}(\epsilon)\!-\!f_{\bar \ell}(\epsilon) \right\} \Big]
\label{It2}
\end{eqnarray}
where the first term in the right-hand side represents the particle flux
from the lead $\ell$ and the second is the flux from the opposite-side lead ${\bar\ell}$.
Here, $T^F_\ell(t,t;\epsilon)$ and $T^R_\ell(t,t;\epsilon)$ are transport coefficients
representing tunneling of holes (electrons) in the region of positive (negative) energies.
The function $T^F_\ell(t,t;\epsilon)$ is a diagonal part of,
\begin{eqnarray}
T^F_\ell(t,t_0;\epsilon) &=&4\pi i \int_{-\infty}^\infty d\tau e^{i p_\ell(t)}
\Sigma_{\bar\ell}^C(t,\tau) e^{-i p_\ell(\tau)}\nonumber\\
&& n_\ell(\tau,t_0;\epsilon)
\label{TF}
\end{eqnarray}
which depends on both sides of the tunneling barriers and can be considered as
the transmission function from the lead $\ell$ to the other side ${\bar \ell}$.
In writing $T_\ell^F$, we adopt an approximated form for its further use
by substituting $\Sigma_{\bar\ell}^C$ instead of  $2 \Sigma^R_{\bar\ell}$.
This approximation corresponds to  the wide-band limit.\cite{jauho}
With this transmission function alone, the expression of Eq. (\ref{It2})
is consistent with the well-known
Laudauer formula.\cite{landauer}
However, the additional function $T^R_\ell$ gives rise to a deviated form from the formula;
$T^R_\ell$ is given by,
\begin{eqnarray}
T^R_\ell(t,t_0;\epsilon) &=&\!8\pi i \left( i\hbar\frac{\partial}{\partial t}-\Delta_0\right)
n_\ell(t,t_0;\epsilon).
\label{TR}
\end{eqnarray}
According to this, $T^R_\ell(t,t_0;\epsilon)$ is independent
of the other-side tunneling barrier $\bar\ell$ contrast to $T^F_\ell$ implying
not a transmission function and,
moreover, it does not appear in the flux from the other-side lead ${\bar \ell}$.
These facts are likely to interpret the function as a transport coefficient describing
the electron or hole flux supplied by the metallic island. This becomes more apparent if
one compares $T^R_\ell$ with Eq. (\ref{Q2}) where the flux by $T^R_\ell$ is equal to
the decrease rate of charges in the island.
This flux has no a stationary component to be $\Re T^R_\ell(t,t;\epsilon)=0$, by which
our expression successfully recovers the Landauer formula in a static condition.

On the other hand, the charge noise is rearranged to be,
\begin{eqnarray}
S_Q(t,t_0) \!&=&\! \sum_{\ell\ell'} S_Q^{\ell\ell'}(t,t_0) \nonumber\\
S_Q^{\ell\ell'}(t,t_0) \!&=&\! 
4 e^2 \int_{-\infty}^{\infty}\!d\epsilon_1\!d\epsilon_2~
n_\ell(t,t_0;\epsilon_1)  n_{\ell'}(t_0,t;\epsilon_2)\nonumber\\
&&     \left\{f_\ell(\epsilon_1) f_{{\bar\ell}}^+(\epsilon_2) 
                      +f_\ell^+(\epsilon_1) f_{{\bar\ell}}(\epsilon_2) \right\},
\label{SQ2}
\end{eqnarray}
which is the sum of all possible electron-hole correlations tunneled from  both leads.

The current noise in time-dependent cases is found to have a complicated form due to
the fluctuations arising from various origins and, therefore,
we show it under the wide-band approximation separating into three contributions.
The self correlation of the tunneling current at the lead $\ell$ is arranged as,
\begin{eqnarray}
S_{\ell \ell}(t,t_0) = S_v^\ell(t,t_0)+ S_r^\ell(t,t_0)+
\frac{\partial^2}{\partial t\partial t_0}
S_Q^{\ell\ell}(t,t_0)
\label{SI2}
\end{eqnarray}
while the correlation between the different-side tunneling currents
can be obtained from Eq. (\ref{Scon}) together with the above one.
Here, the first term reads,
\begin{eqnarray}
S_v^\ell(t,t_0) =
\frac{1}{e^2R_K^2} \int_{-\infty}^{\infty} d\epsilon_1d\epsilon_2
  L_{\ell }(t,t_0;\epsilon_1)
                   n_{\bar\ell}(t_0,t;\epsilon_2) \nonumber\\
 \left\{f_{\ell}(\epsilon_1) f_{\bar\ell}^+(\epsilon_2)
\!+\!f_{\ell}^+(\epsilon_1) f_{\bar\ell}(\epsilon_2) \right\} 
\end{eqnarray}
where 
$L_{\ell }(t,t_0;\epsilon)$ is defined by,
\begin{eqnarray}
L_\ell(t,t_0;\epsilon) =8\pi i e^{ip_\ell(t)-ip_\ell(t_0)} 
e^{-i\epsilon(t-t_0)/\hbar} {\tilde \Sigma}_\ell^C(\epsilon).
\end{eqnarray}
Actually, in a static condition this term is proportional to the transmission coefficient $T^F$
and represents the noise arising from backward tunneling after electrons and holes starting
from different leads, respectively, tunnel into the island simultaneously.
As a result, the term is not correlated by the Coulomb interaction in the island,
by which it reproduces the results of the co-tunneling theory.\cite{averin1,sukhor}
The second term of Eq. (\ref{SI2}) represents
the correlation between the real tunneling currents;
\begin{eqnarray}
S_r^\ell(t,t_0) = \frac{2}{e^2 R_K^2}\Re
\int_{-\infty}^{\infty} d\epsilon_1\!d\epsilon_2 \Big[ ~~~~~~~~~~~~~~~~~~~~~~~~~~~~~~~\nonumber\\ 
T^R_{\ell }(t,t_0;\epsilon_1) T^F_{\bar\ell}(t_0,t;\epsilon_2)
 \left\{f_{\ell}(\epsilon_1) f_{\bar\ell}^+(\epsilon_2)
\!+\!f_{\ell}^+(\epsilon_1) f_{\bar\ell}(\epsilon_2) \right\} \nonumber\\
-T^R_{\ell }(t,t_0;\epsilon_1)
                  T^F_{\ell}(t_0,t;\epsilon_2)
 \left\{f_{\ell}(\epsilon_1) f_{\ell}^+(\epsilon_2)
\!+\!f_{\ell}^+(\epsilon_1) f_{\ell}(\epsilon_2) \right\}\nonumber\\ 
+\sum_{\ell'\ell''} (\delta_{\ell',\ell''}\!-\!\frac{1}{2}) T^F_{\ell' }(t,t_0;\epsilon_1)
T^F_{\ell''}(t_0,t;\epsilon_2) ~~~~~~~~~~~~~~\nonumber\\
\left\{f_{\ell'}(\epsilon_1) f_{\ell''}^+(\epsilon_2)
\!+\!f_{\ell'}^+(\epsilon_1) f_{\ell''}(\epsilon_2)\! \right\}\! \Big ],~~~~~
\label{Srl}
\end{eqnarray}
where each transport coefficient of $T^F$ and $T^R$ represents the corresponding current
and, thereby, the product of them describes their correlation.
In a static case, the function $T^R_\ell$ has no a real value, so that this term recovers
the well-known form that represents the correlated tunneling processes via the Coulomb 
interaction.  By recalling the term $S_r^\ell$ to the self correlation,
some of the correlations between $T^R$ and $T^F$  are missing or additionally incorporated 
in Eq. (\ref{Srl}).
This is a consequence of the last term which emphasizes the role
of the charge fluctuation in the island.
In terms of the noise spectrum, this term is written as $\omega^2 S_Q^{\ell\ell}(\omega)$.

\subsection{Adiabatic limit}

For sufficiently small frequencies of driving fields, it is expected that
the time-dependent fields just vary the chemical potentials adiabatically, so that
formal expressions are the same as those in static problems.
This is really the case if one expands phases in the self-energy as $p_\ell(t_1)=p_\ell(t_0)+{\dot p}_\ell(t_0)(t_1-t_0)+\ldots$ and neglects higher-order terms;
\begin{eqnarray}
\Xi_\ell(t_1,t_2) &=& e^{i p_\ell(t_1)} \Sigma_{\ell}(t_1,t_2) e^{-i p_\ell(t_2)}\nonumber\\
&\sim& e^{i {\dot p}_\ell(t_0)(t_1-t_2) } \Sigma_{\ell}(t_1,t_2).
\label{Xiapp}
\end{eqnarray}
Then, this changes simply the chemical potential in Eq. (\ref{Sigma0}) as;
\begin{eqnarray}
\mu_\ell^0 \rightarrow \mu_\ell(t_0)\!=\!\mu_\ell^0\!-\!ev^b_\ell(t_0)
\!-\!(\delta_{\ell,D}\!-\!C_D/C_\Sigma)eV_D^0
\end{eqnarray}
where we separate the potential difference across a tunneling barrier $\ell$ into
an alternating part of $v_\ell^b(t)$ and a direct part with a static external voltage $V_D^0$ of
$V_D(t)$.  Here, we use the time argument $t_0$ to clarify the fact that
the chemical potentials in noise calculations depend on the reference time $t_0$ rather
than another time argument $t$.

Under the adiabatic approximation,
the effective density of states and the transport coefficients are simplified to;
\begin{eqnarray}
n_\ell(t,t_0;\epsilon) &=& 
\frac{i}{4 \pi }
{\tilde\Sigma}_{\ell}^C(\epsilon)
\mid\!G^A(\epsilon)\!\mid^2 e^{-i \epsilon(t-t_0)/\hbar}, \nonumber\\
T^F_\ell(t,t_0;\epsilon) &=&
4\pi i {\tilde\Sigma}_{\bar\ell}^C(\epsilon) n_\ell(t,t_0;\epsilon),\nonumber\\
L_\ell(t,t_0;\epsilon) &=&8\pi i {\tilde \Sigma}_\ell^C(\epsilon) e^{-i\epsilon(t-t_0)/\hbar}
\label{Tortho}
\end{eqnarray}
and a spectral form of $G^A(t,t_0)$ is given by,
\begin{eqnarray}
\tilde G^A(E) = \frac{1}{E-\Delta_0-\tilde\Sigma^A(E)}.
\label{GA0}
\end{eqnarray}
Substituting the above expressions into Eqs. (\ref{Q2}), (\ref{It2}), (\ref{SQ2}), and (\ref{SI2}),
one can find that static results
in the work of Ref. \cite{utsumi} are exactly recovered.
The function $T^R_\ell$ has no effects on results because its value is  imaginary
in the approximation.

It is interesting to find the boundary within which the adiabatic expressions are valid.
To do this, we propose the transmission function $T^F(t,t_0;\epsilon)$ in Eq. (\ref{TF}) as a precursor.
Then, expanding phases in the self-energy as in Eq. (\ref{Xiapp}) and 
requiring the zero\_th order term of $T^F$ much larger than 
the largest non-adiabatic contribution (the first-order term),
one obtain a criterion for the valid adiabatic expressions as,
\begin{eqnarray}
\frac{(\hbar\omega_A)^2}{(\hbar\omega_A)^2+4\gamma_0^2 }
\frac{ \mid \sum_{\ell}e v^b_\ell(t)/R_\ell\mid}{
\mid \sum_{\ell}[ \Delta_0-\mu_\ell(t)]/R_\ell\mid}
\ll 1
\label{criterion}
\end{eqnarray}
where the absolute sign means a root-mean-square value of a time-dependent function,
$\omega_A$ is a frequency of external perturbation,
and a total tunneling rate $\gamma_0 \equiv \Im \Sigma^A(\Delta_0)$.
According to this criterion, the adiabatic approximation is well hold
under the case of a smaller applied frequency $\omega_A$ than 
the total tunneling rate $\gamma_0$ together with 
relatively weak amplitudes of perturbations, as usually expected.
However, in general, it also depends on temperature, capacitances of the tunneling barriers,
as well as the excitation energy $\Delta_0$.

\subsection{Orthodox results }

Under the adiabatic regime we further approximate the formalism by assuming 
very opaque tunneling barriers.
Then, $\mid\! \tilde G^A(\epsilon)\!\mid^2$ behaves a nearly $\delta$-function,
reflecting a long life time of charged states.
This makes possible the integration in the formalism as,
\begin{eqnarray}
\int_{-\infty}^{\infty}d\epsilon F(\epsilon) \mid\!G^A(\epsilon)\!\mid^2 =
\frac{\pi }{\gamma_0} F(\Delta_0) \nonumber\\
\int_{-\infty}^{\infty}d\epsilon F(\epsilon) \mid\!G^A(\epsilon)\!\mid^2
 \mid\!G^A(\epsilon-\omega)\!\mid^2 = \nonumber\\
\frac{\pi }{\gamma_0(4\gamma_0^2+\omega^2)}[ F(\Delta_0)+F(\Delta_0+\omega)]
\end{eqnarray}
where we neglect a renormalization effect on the excitation energy $\Delta_0$.
As a consequence, results become identical to those based on the orthodox theory.\cite{johan,korotkov}

\section{ Applications of formalism }

As applications of our formalism, we now examine the performance of rf SETs numerically.
For this, we consider detailed geometry of the circuit which is characterized by 
the impedances ($Z_G$ and $Z_D)$ and the voltage sources ($V_D$ and $V_G$) in Fig. \ref{fig1}.
According to measuring processes of signals, there are two kinds of SETs;
reflected and transmitted types.
In this work we assume the reflected type of rf SETs where a detector measures reflected signals
from a SET.\cite{aassime,schoel,buehler}
Applications  to the transmitted type\cite{fujisawa,cheong}
is basically identical with slightly modified electrical environments.

Then, an equivalent circuit for the rf SET is given
by writing the impedances  as
\begin{eqnarray}
Z_G(\omega) &=&  0 \nonumber\\
Z_D(\omega) &=&  R_0 \frac{1+i\tilde{\omega} Q_F}{1-\tilde{\omega}^2 +i\tilde{\omega}/Q_F}
\label{rfZ}
\end{eqnarray}
where $R_0$ is the coaxial cable impedance (typically 50$\Omega$),
$\tilde{w}=\omega/\omega_R$ is a normalized frequency with a resonant frequency $\omega_R$
of the tank circuit, and $Q_F$ is its quality factor.
%
We model microwaves propagating the coaxial cable
as a sinusoidal form, $v^-(t) = v_{in} cos(\omega_A t)$
with amplitude and angular frequency $v_{in}$ and $\omega_A$, respectively, and also consider
a static voltage $V_D^0$ provided by a bias tee.
Then, the equivalent voltage source of $V_D(t)$ is given by,
in its Fourier component,
\begin{eqnarray}
\tilde{V}_D(\omega) = \frac{Z_D(\omega)}{R_0} \frac{2 v^-(\omega)}{1+i \tilde{\omega} Q_F}
+\delta(\omega)V_D^0
\end{eqnarray}
or in a real-time space,
\begin{eqnarray}
V_D(t) &=& 2 \eta_0(\omega_A)v_{in}\cos(\omega_A t+\varphi(\omega_A))
+V_D^0
\end{eqnarray}
with  the phase and amplitudes defined by
\begin{eqnarray}
\eta_0(\omega) e^{i\varphi(\omega)} &=&  \frac{1}{ 1-\tilde\omega^2+i\tilde\omega/Q_F }.
\end{eqnarray}
A gate voltage $V_G(t)$ represents a signal to be measured, for instance,
the voltage induced by time-varied charges of qubits.
If the signal is slowly varied in time, it can be treated in the adiabatic way
like the change of a static voltage $V_G^0$, which in turn gives rise to
the modulation of the excitation energy $\Delta_0$ adiabatically.
In the followings we assume this case where  $V_G(t)$ is no longer a voltage source, i.e.,$V_G(t)=0$,
but a parameter for the excitation energy $\Delta_0$.
However, in general it should be treated time-dependent fields for considerations
of gate charges varied rapidly.

From the voltage sources, the potential differences across the tunneling barriers
from Eq. (\ref{vrb}) become
\begin{eqnarray}
\left(
\begin{array}{c}
{v}^b_D(t)\\
{v}^b_S(t)\\
\end{array}
\right)
=
\left(
\begin{array}{cc}
 1-\frac{C_D}{C_\Sigma} &-\frac{C_G}{C_\Sigma} \\
-\frac{C_D}{C_\Sigma}     &-\frac{C_G}{C_\Sigma}\\
\end{array}
\right)
\left(
\begin{array}{c}
V_D(t)-V_D^t(t)\\
V_G(t)-V_G^t(t)\\
\end{array}
\right)
\label{vdreal}
\end{eqnarray}
with $V_G(t)=0$.
Here,
the induced voltages from the tunneling currents are calculated from Eq. (\ref{Vteff}) as,
\begin{eqnarray}
V_D^t(t) \!&=&\!\frac{1}{T_A}\int_0^{T_A}d\tau
I_D^t(\tau) \Big[R_0 \!+\!2\Re Z_D(\omega_A) e^{i\omega_A (t\!-\!\tau)} \Big]
\nonumber\\
V_G^t(t) \!&=&\!-\frac{1}{T_A}\int_0^{T_A} d\tau
[ I_D^t(\tau)\!+\!I_S^t(\tau) ] 2\Im \frac{e^{i\omega_A (t\!-\!\tau)}}{i\omega_A C_G}, 
\end{eqnarray}
where, due to resonant properties of the tank circuit,
the lowest two harmonics of the tunneling currents are taken into account and $T_A=2\pi/\omega_A$ is
a period of the microwave.

As for output signals, we consider 
a reflected voltage $v^+(t)$ from the SET;
\begin{eqnarray}
v^+(t) &=& v_{in} \cos[\omega_A t-2\varphi(\omega_A)]-R_0 I_0
\nonumber\\
&+&X \cos(\omega_A t)+Y\sin(\omega_A t)\nonumber\\
&+& {\rm higher~harmonics}.
\end{eqnarray}
Here, we separate a pure reflected component of
the microwave (the first term in the right-hand side)
from its responses to the single-electron transistor.
Each term depending on $I_0$, $X$, and $Y$ is originated from
the tunneling current,
\begin{eqnarray}
I^t(t) \!&=&\! \left(1-\frac{C_D}{C_\Sigma}\right )I_D^t(t)-\frac{C_D}{C_\Sigma} I_S^t(t) \nonumber\\
\!&=&\! I_0 \!+\! I_1 \cos(\omega_A t)\!+\! I_1'\sin(\omega_A t)\nonumber\\
&\!+\!&{\rm higher~ harmonics},
\label{totI}
\end{eqnarray}
and their Fourier components are related to each other as,
\begin{eqnarray}
X &=&-\frac{R_0}{\sqrt{(1-{\tilde\omega}^2)^2+{\tilde\omega}^2/Q_F^2}} I_1 \nonumber\\
Y &=&-\frac{R_0}{\sqrt{(1-{\tilde\omega}^2)^2+{\tilde\omega}^2/Q_F^2}} I_1'.
\label{X}
\end{eqnarray}

To model a detector we consider the average of observables multiplied by $\cos \omega t$
or $\sin\omega t$ over time.\cite{turin,roschier}
Especially, we focus on a homodyne detector measuring the amplitude $X$ 
obtained from the reflected voltage $v^+(t)$.
As indicated in Ref. \cite{turin}, since the amplitude $X$ is usually much larger than $Y$
for  $Q_F\gg 1$ implying a small reactance of tunneling barriers, it is
a good approximation to express the performance of the rf SET only in terms of $X$.
After many numerical simulations we find that this is also the case for our system.

Then, the noise associated with $X$ is derived as,
\begin{eqnarray}
S_{X\!X}(\omega) \!=\!
\frac{ 2R_0^2}{(1-{\tilde\omega}^2)^2+{\tilde\omega}^2/Q_F^2} S_I(\omega)
\label{Sxx}
\end{eqnarray}
with  noise of a tunneling current $S_I(\omega)$,
\begin{eqnarray}
S_I(\omega) &\!=\!& 
\sum_{\ell\ell'}(\delta_{\ell,D}\!-\!\frac{C_D}{C_\Sigma})
(\delta_{\ell',D}\!-\!\frac{C_D}{C_\Sigma}) \nonumber\\
&& \langle\!\langle S_{\ell\ell'}(t+t_0,t_0)
[ \cos\{2\omega t_0+2\varphi(\omega)\} \!+\!\cos \omega t  ] \rangle\!\rangle.
\label{SIexp}
\end{eqnarray}
It is noted that the current noise of $S_I(\omega)$ is determined from the average weighted
with a factor  $\cos\{2\omega t_0+2\varphi(\omega)\}$ as well as $\cos \omega t$.
This is a consequence of modeling the homodyne detector.
In static cases,
the average with the former gives zero
because the current fluctuation of $S_{\ell\ell'}(t+t_0,t_0)$ is invariant under time-translation,
and then $S_I$ just represents  noise of the total current $I^t$ of Eq. (\ref{totI}).
However, in general since the invariance is no longer valid under time-dependent conditions as
easily checked in Eq. (\ref{SI2}), the average  has a finite value and,
as a result, the noise of $X$ contains additional contributions
at every multiples of the half-frequency $\omega_A/2$.

The zero-frequency approximations correspond to setting of $\omega=0$ in
$\cos \omega t$ for considering a small frequency $\omega$
and then our expression of Eq. (\ref{SIexp}) becomes similar to Eq. (30) of Ref. \cite{turin}.

Our numerical examinations are fulfilled under several limitations due to
the two-states approximation.
So, the range of parameters for operating points such as the excitation energies
$\Delta_0$ (or gate charges $q_0$),
the amplitudes $v_{in}$ and frequencies $\omega_A$ of microwaves,
 and the DC bias voltage $V_D^0$ should be restricted not to
occupy higher or lower charged states.
For this we consider the excitation energies in the range of
$-E_C\leq \Delta_0\leq E_C$ (or $0\leq q_0\leq e$)
and microwave energy $\hbar\omega_A$ much less than $E_C$.
When the frequency of a microwave is tunned to be resonant
to the tank circuit (i.e., $\omega_A=\omega_R$, which is also assumed
for our numerical calculations),
the maximum amplitude of $2Q_F v_{in}$ is delivered to the SET.
Then, not to excite other charged states,
the applied amplitudes should satisfy the inequality of
\begin{eqnarray}
\mid eV_D^0\mid\!+\!2 Q_F e v_{in} &\!\le\!&
{\rm min}\!\left\{ \frac{C_\Sigma}{C_D},\frac{C_\Sigma}{C_\Sigma\!-\!C_D}\right\}
\nonumber\\
&& 2E_C(1\!-\!\left| \frac{q_0}{e}\!-\!\frac{1}{2}\right| )
\end{eqnarray}
which can be derived from simple electrostatic consideration.
According to this, energy provided by the drain voltages (the left hand term in the above equation)
is restricted to be approximately less than $E_C$ and $2 E_C$ at completely blockade points ($q_0=0,e$)
and a degenerate point ($q_0=e/2$), respectively.

Hereafter
we use local units for calculated results and system parameters.
We display all quantities in units of  charging energy $E_C$, but in units of an ohm for
resistances.
In other words, thermal energy $k_BT$, excitation energy $\Delta_0$,
amplitude $ev_{in}$ and frequency $\hbar\omega_A$ of microwaves,
and DC bias voltages $eV_D^0$ are measured in units of $E_C$ as well as
currents are measured in units of $e E_C/\hbar$ and capacitances in units of $e^2/E_C$.
In these units, the total capacitance becomes $C_\Sigma=0.5$.
So, the situation such as $C_D=C_S=0.2$ in the followings
means a small gate capacitance and, therefore, gives  rise to
somewhat symmetric distribution of potential differences
across the tunneling barrier from Eq. (\ref{vdreal})
while the case of $C_D=C_S=0.01$ represents asymmetric distribution.
Our numerical calculations are performed for $R_0=50\Omega$, a zero static voltage $V_D^0=0$,
and a temperature of $k_BT=0.02 E_C$ (unless mentioned otherwise).
Even though the dependence of the SET performance  on  temperature and finite static voltages are also
interesting,\cite{schoel,turin} we omit it for simplicity.

\subsection{ Environmental effects }

\begin{figure}
\centering
\includegraphics[width=0.4\textwidth]{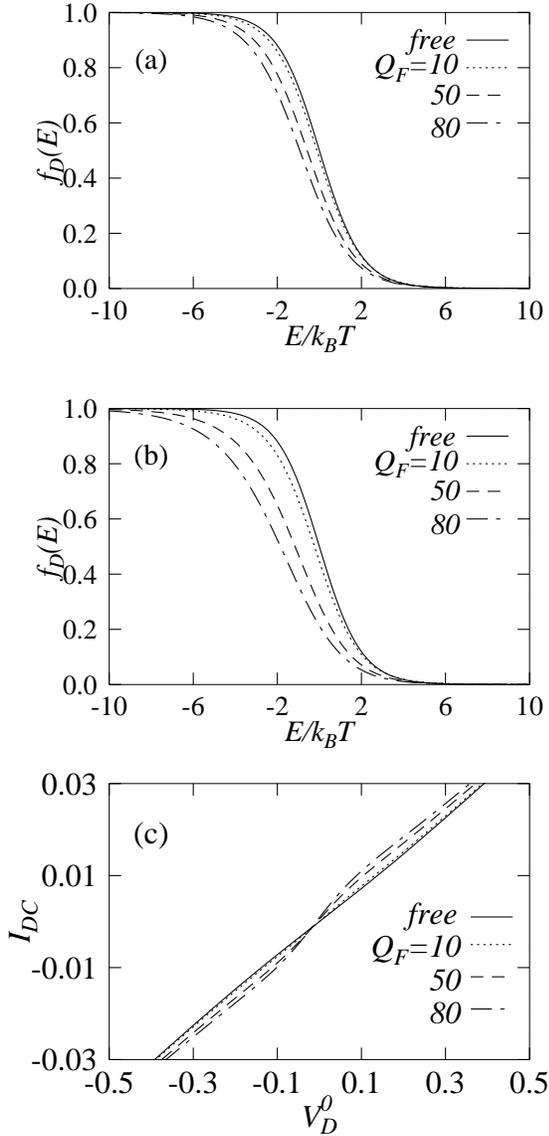}
\caption{
We show the Fermi-Dirac distributions at the drain for two electrical environments.
In (a), we choose a resonant frequency of $\hbar\omega_R=0.04$ at a temperature
of $k_B T=0.02$ while in (b) $\hbar\omega_R=0.08$ and  $k_B T=0.02$.
For various quality factors of $Q_F=10$ (dotted line), $50$ (dashed), $80$ (dot-dashed),
results are compared with that in the case of the free environment (solid).
In (c) DC tunneling currents are calculated as a function of DC drain voltage $V_D^0$
with parameters in (b) at $q_0=e/2$. Here we  use
$R_D=R_S=13.1k\Omega$ $(\alpha_0=0.1)$
under a nearly symmetric configuration of barriers, $C_D=C_S=0.2$.
}
\label{envFig}
\end{figure}

Firstly we discuss effects of the electrical environments.
For the given electric environment of Eq. (\ref{rfZ}),
its effects are manifested into the self-energy of $\Sigma_\ell(E)$.
In general, the presence of electrical environments broaden the self-energies
and one can identify this via the Fermi-Dirac distribution functions of Eq. (\ref{FD}).
In Fig. (\ref{envFig}), we show the Fermi-Dirac distribution functions for two cases
of resonance frequencies for various quality factors, in which
the resonance frequency is comparable to thermal energy in $(a)$ and
much larger than it in $(b)$.
Characteristic behavior is that the distribution function becomes more depleted
around the chemical potential $(\mu^0_\ell =0)$
as the quality factors and the resonance frequencies increase.
This is resulted from the energy-emitting and absorbing spectrum of the environment.
As indicated in Ref. \cite{grabert} a small quality factor means rather rapid damping
in electrical dynamics of the environment or Ohmic behavior,
in which  the spectrum  has a peak around a zero irrelevant of the resonance frequency.
Whereas, a large quality factor leads to the environment with a single mode case where
quantized energy equal to $\hbar\omega_R$ is incorporated into the spectrum and
produce peaks at every multiple of $\hbar\omega_R$.
Thus, in the case of the large quality factor, tunneling is mediated to the environment
by emitting or absorbing energy quanta of $\hbar\omega_R$. As a result, if  thermal energy $k_B T$ is less than it
and cannot  excite the environment,
the energy quanta should be provided externally, which in turn reduces
the effective number of particles for tunneling.

To emphasize the effect of the depletion in the particle distributions we show 
calculated tunneling currents for static cases in Fig. \ref{envFig}-$(c)$
using parameters in $(b)$.
For a small quality factor ($Q_F=10$), tunneling currents are very similar to that
for the free environment.
However, for a large value of $Q_F=80$, noticeable differences are found.
This difference becomes more enhanced for larger resonant frequencies over the thermal energy
and more asymmetric geometry of tunneling barriers like small capacitances of $C_D$ and
$C_S$ compared to $C_G$.

\subsection{ Effects of photon-assisted tunneling}

Next we discuss effects of time-dependent perturbations 
on transport by comparing results calculated from the exact and adiabatic expressions.
For this we consider sufficiently large amplitudes
and high frequencies of a microwave compared to a temperature as well as asymmetric
geometry of barriers.
Calculations are performed for two cases of tunneling resistances; one is
 large resistance of $R_D\!=\!R_S\!=\! 655k\Omega~ (\alpha_0\!=\!0.002)$ and
the other is a small one, $13.1k\Omega~ (\alpha_0\!=\!0.1)$.
So, the case of the former is believed to be dominated by sequential tunneling in transport
while with the small resistance both
co-tunneling and sequential processes are expected to be contributed.

\begin{figure}
\centering
\includegraphics[width=0.4\textwidth]{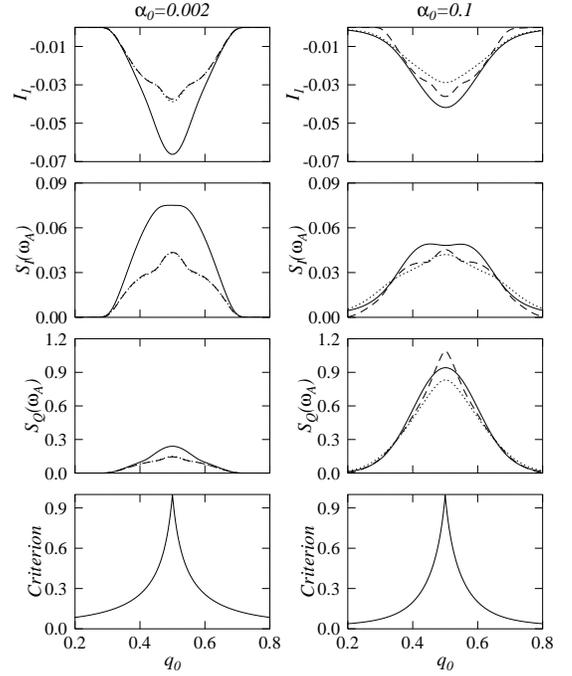}
\caption{
We plot the first harmonics of a tunneling current (in units of $E_C/eR_T$),
current and charge noises (in units of $E_C/R_T$ and $e^4 R_T/E_C$, respectively),
and the criterion function as a function of gate charge $q_0$
for two sets of parameters.
In the left column, tunneling resistances are chosen to be $R_D=R_S=655k\Omega$ ($\alpha_0=0.002$) and
a frequency of a microwave $\hbar\omega_A\!=\!\hbar\omega_R=0.01$ while
in the right column $R_D=R_S=13.1k\Omega$ ($\alpha_0=0.1$) and $\hbar\omega_A= 0.2$ are used.
We compare results calculated based on the exact, adiabatic, and orthodox formula
by displaying them in solid, dotted and dashed lines, respectively.
Other parameters are $k_BT=0.01$, $C_D=C_S=0.1$, $Q_F=6$, and $2Q_F ev_{in}=0.5$.}
\label{adiaFig}
\end{figure}
Fig. \ref{adiaFig} shows calculated results of the currents and noises for the above two
cases as a function of gate charge; for the smaller (large) in the left (right) column.
In each column we additionally distinguish results depending on the formalism used;
the exact (solid), adiabatic (dotted), orthodox (dashed) formalism, respectively.
Thus, the differences between the exact and adiabatic results represent the existence
of photon-assisted tunneling while those between adiabatic and orthodox results just emphasize
effects of co-tunneling in the classical limit (for the small resistance the orthodox theory
is known to be invalid, nevertheless we also show corresponding results just for comparison).

\begin{figure}
\centering
\includegraphics[width=0.4\textwidth]{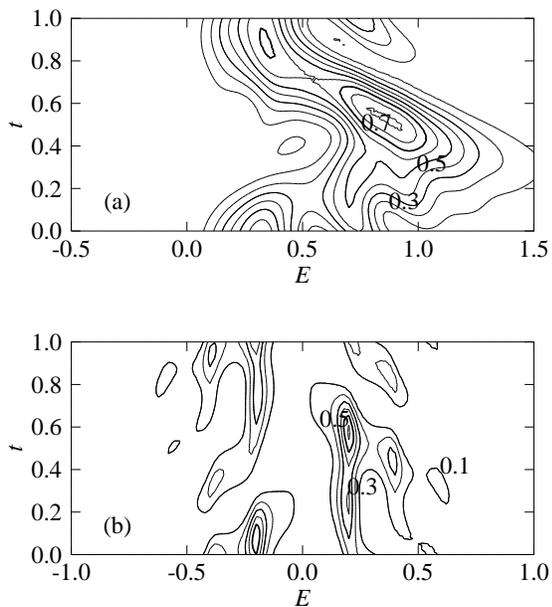}
\caption{
We plot the transmission function $T^F_D(t,t;E)$ at the drain with parameters
used in Fig. (\ref{adiaFig})-(b).
For gate charges of $q_0=0.2e$ and $0.5e$, its contour plot
is depicted in $(a)$ and $(b)$, respectively. Here, a time $t$ is measured in units of
a period of a microwave $T_A$. }
\label{tfFig}
\end{figure}
From the calculated results, one can see that
the system is easily driven into the non-adiabatic regime
around the degeneracy point $q_0=e/2$ in both cases of tunneling resistances
implying the fact that tunneling occurs in photon-assisted ways and,
there, tunneling currents as well as noises are largely enhanced relative to the adiabatic results.
This tendency is more apparent in the case of the large resistance.
However, apart far from the degeneracy point, effects of time-dependent perturbations are
reduced to classical ones to give nearly identical results independently of the formalism used.
In other words, the strength of the photon-assisted tunneling becomes weak far from 
the degeneracy point.

One of possible explanations for the different strength of the photon-assisted tunneling
depending on gate charges $q_0$ and tunneling resistances is
the resolution of photon energies $\hbar\omega_A$ seen by particles.
As inferred from Eq. (\ref{GA0}) since particles in the island is decayed with
a rate of $\gamma_0=\Im \Sigma^A(\Delta_0)$, 
its dwelling time can be regarded to be inversely proportional to the rate.
Along this aspect, the left-hand term  in Eq. (\ref{criterion}) can be interpreted as
rough estimator of the resolution for energy quanta $\hbar\omega_A$.
We plot it in the bottom panel of Fig. \ref{adiaFig} for both cases of the tunneling resistances.
According to the figure, when the value is less than about $0.1$,
the calculations show nearly identical behavior between the adiabatic and exact results 
implying poor resolution for energy quanta.
This fact is also confirmed by studying the transmission function $T^F$ of Eq. (\ref{TF}).
Fig. \ref{tfFig} is a contour plot of the transmission function
in the energy-time space for two different points of gate charges at which
the resolution is predicted to be poor (at $q_0=0.2e$) and good (at $q_0=0.5e$)
in Fig. \ref{adiaFig}-$(b)$.
As shown in the figure, their time and energy dependences are strikingly complex.
However, one can recognize the distinct feature different from each other.
In $(a)$ for a given time the height of the transmission function is changed continuously
as a function of energy while  in $(b)$ it exhibits rather discrete behavior.
Actually in the case of $(b)$ the distance between successive maximums is equal to 
photon energy of $\hbar\omega_A$, which is well resolved enough to see photon-side bands.


\begin{figure}
\centering
\includegraphics[width=0.4\textwidth]{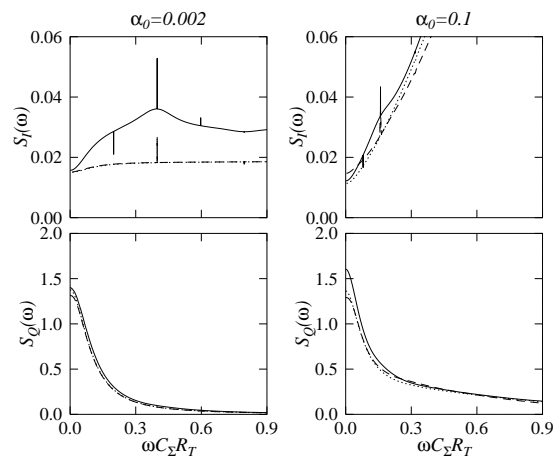}
\caption{
Under the same condition as those of Fig. \ref{adiaFig}, we plot the frequency dependence of
current (top) and charge (bottom) noises for $\alpha_0=0.002$ (left) and $0.1$ (right panel), 
respectively. In each figure, results by the exact, adiabatic, and orthodox formula
are represented by solid, dotted, and dashed lines, respectively,
at a gate charge $q_0=0.4$ (current and charge noises are measured in units of 
$E_C/R_T$ and $e^4 R_T/E_C$, respectively).
Sharp peaks correspond to an infinite averaging time $T$ in Eq. (\ref{daverage}),
otherwise smeared ones may be obtained.
}
\label{wdepFig}
\end{figure}
For the frequency dependences of the current and charge noises, we examine them
at a gate charge $q_0=0.4e$ of Fig. \ref{adiaFig} and show results
in  Fig. \ref{wdepFig} in the range of low frequencies.
In the current noises, the differences at a frequency $\omega=\omega_A$
between the adiabatic and exact results are found to be retained 
in the whole range of frequencies accompanying some
structured behavior in both tunneling resistances.
A straightforward explanation for these differences is complicated by
various correlations among transport coefficients
of $n_\ell$, $T^F_\ell$, $T^R_\ell$, and $L_\ell$
in Eq. (\ref{SI2}) where they all are found to play roles somewhat importantly.
As for peaks located at multiple of a frequency $\omega_A/2$,
they are apparently arisen from time-dependent properties of external perturbations
as noted in Eq. (\ref{SIexp}).
Usually, the largest peak is found at $\omega=\omega_A$, however,
in the exact results peaks at $\omega=\omega_A/2$ and $3\omega_A/2$ are
also appreciable contrary to the adiabatic cases.
This means that in photon-assisted ways the correlations between different harmonics of
the above coefficients remain large while they are unimportant in the adiabatic limit.
In the bottom of the figure, we show the charge noises as a function of frequency.
Due to a large frequency scale we do not distinguish their differences clearly.
However, in term of $\omega^2 S_Q(\omega)$, we find that each result shows
an appreciable deviation from the others.
Actually, the function of $\omega^2 S_Q(\omega)$ is related to the current noises
because the last term of Eq. (\ref{SI2}) is approximately equal to it
neglecting geometrical factors between capacitances.
By comparing it with the current noises, we find that
a large portion of the difference in the current noise 
can be associated with those in the charge noises.

For high frequencies, the noises are found to be nearly identical to
equilibrium noises calculated with no external perturbations. Namely,
the Johnson-Nyquist noises are dominant.

\subsection{ Charge sensitivity}

The sensitivity to gate charges is one of estimators for the performance
of a rf SET as an electrometer. 
If one measures the amplitude $X$ assuming a homodyne detector,
the charge sensitivity is calculated as,\cite{averin,turin}
\begin{eqnarray}
\delta q = \frac{\sqrt{S_{XX}(\omega_A)}}{\mid d X/d q_0\mid}
= \frac{\sqrt{ 2S_{I}(\omega_A)}}{\mid d I_1/d q_0\mid},
\label{dq}
\end{eqnarray}
showing the dependence  on the current noises $S_I$ and the response function $dI_1/dq_0$.
\begin{figure}
\centering
\includegraphics[width=0.4\textwidth]{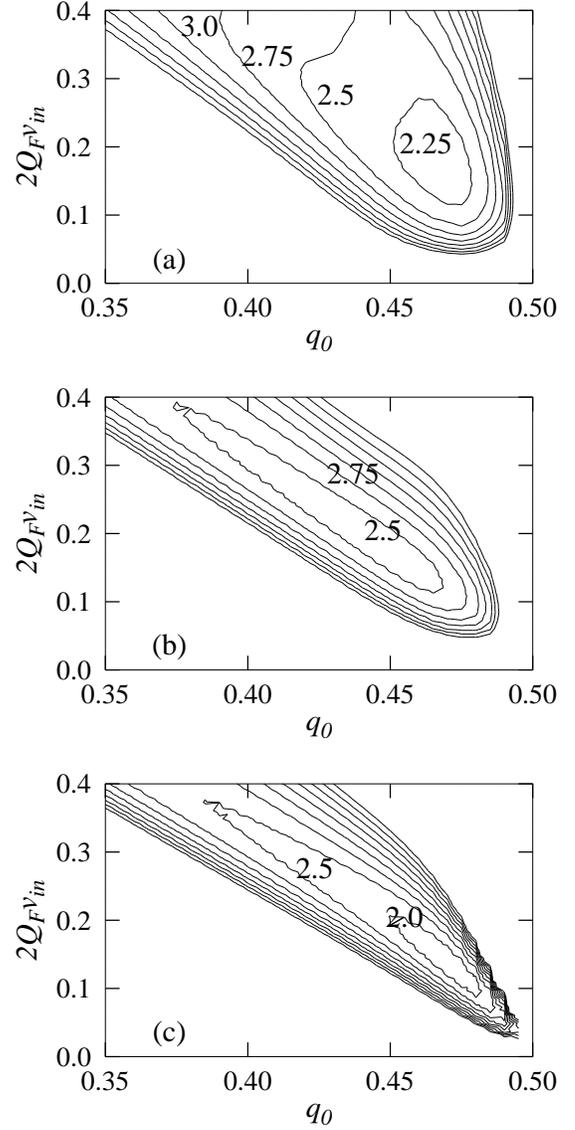}
\caption{
Contour plots of the charge sensitivity $\delta q$ (in units of $e\sqrt{\hbar/E_C}$)
in the space of rf amplitudes and gate charges are shown.
The used parameters are $\hbar\omega=0.04 E_C$, $Q_F=6$, and $R_D=R_S=100k\Omega$
at a temperature of $k_B T=0.02 E_C$.
}
\label{senFig}
\end{figure}
In Fig. \ref{senFig}  we plot the sensitivity in the $q_0$-$v_{in}$ plane
using the three different formula. 
Aiming at the simulation of Ref. \cite{schoel}, we use similar system parameters
to the experiment except for $C_D=C_S=0.1$ (unfortunately not specified in the experiment).
To clarify calculated results we omit the regions of poor sensitivity, which correspond
to smaller rf-wave amplitudes than the Coulomb blockade thresholds (lower left corner)
and much higher rf-wave amplitude over the threshold with small excitation energies (upper
right corner). 
By comparing Fig. \ref{senFig}-(a) and (b), one can see effects of photon-assistant tunneling
to the sensitivity. That is, the region of  good sensitivity (for example, within $\delta q=2.5$)
is predicted to slowly vary as a function of rf-wave amplitude than in the adiabatic limits,
while it exhibits a rather narrower region as a function of gate charge.
Calculated optimum sensitivities (minimum value of $\delta q$) are also found to have different values,
$\delta q=2.12, 2.38$, and $1.83$ for the exact, adiabatic, and orthodox formalism, respectively,
with a nearly same operating point of $(q_0,2Q_Fv_{in})=(0.47,0.12\sim 0.17)$.
The slightly better optimum-sensitivity in Fig. \ref{senFig}-$(a)$ compared with that in $(b)$
comes from the enhanced value of the response $dI_q/dq_0$ by photon-assisted tunneling.
However, photon-assisted tunneling does not always give the better sensitivity because
it also enhances current noises.
We find that its role for the sensitivity depends on system parameters.
In the orthodox result, the sensitivity is always predicted to be better than results of
the others
because the absence of co-tunneling gives larger values of the response $dI_q/dq_0$.

Using the experimental parameter of $E_C=178\mu eV$,
the calculated optimum sensitivity $\delta q=2.12$ corresponds to $4.1\mu e/\sqrt{{\rm Hz}}$,
which is lower than the measured value of $47\mu e/\sqrt{{\rm Hz}}$ by an order.
This discrepancy between the theory and experiment may be attributed to the preamplifier
noise and local heating of a SET.\cite{turin}

\begin{figure}
\centering
\includegraphics[width=0.4\textwidth]{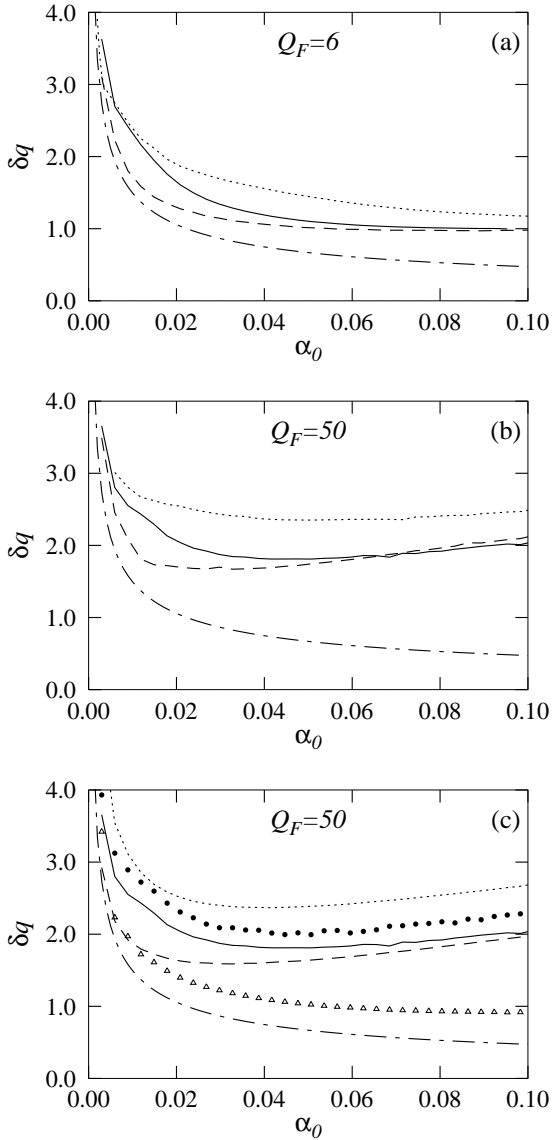}
\caption{
The optimum charge sensitivities are plotted as a function of tunneling resistance for
two quality factors of $Q_F=6$ in $(a)$ and $50$ in $(b)$, respectively, together with
different frequencies of a microwave, $\hbar\omega_A=0.1$ (dotted),
$0.04$ (solid), and $0.005$ (dashed).
The dot-dashed lines denote the analytical result of Eq. (\ref{analq}).
Here, we assume $R_S=R_D$, $C_D=C_S=0.1$, and $k_B T=0.02$.
In (c) we re-plot the result of (b) represented by the solid line,  and examine 
its change without each calculational factor one by one; adiabatic approximation (dotted line),
no electrical environment (square), symmetric geometry of $C_D=C_S=0.24$ (dashed line), and
calculation neglecting a self-consistency (triangle). 
}
\label{senq0Fig}
\end{figure}
In Fig. \ref{senq0Fig}, we show the dependence of the optimum sensitivity
on tunneling resistances for different applied frequencies of microwaves and
quality factors.
Firstly, Figs. \ref{senq0Fig}-$(a)$ and $(b)$ are results for $Q_F=6$ and $50$, respectively.
By comparing different kinds of lines in each figure which means different frequencies
of microwaves at a resonant condition,
one can see that larger frequencies give rise to worse sensitives
in a wide range of tunneling resistance.
Since a large driving frequency corresponds to better-resolved energies of external fields,
associated photon-assisted tunneling is expected to decrease the sensitivity
due to the enhanced current noises.
On the other hand, for a given frequency,
the sensitivity as a function of tunneling resistance
is found to show different behavior depending on the quality factor;
monotonically improved results are obtained for $Q_F=6$
as tunneling resistance decreases while there is optimal resistance
for the best sensitivity for $Q_F=50$.
For easier understanding of our results,
an analytic expression from Ref. \cite{turin},
\begin{eqnarray}
\delta q \sim 2.65 e \sqrt{(R_D\!+\!R_S)C_\Sigma}\sqrt{\frac{k_BT}{2 E_C}}
\label{analq}
\end{eqnarray}
is also plotted with dot-dashed lines in the figure even though
it gives difference values from ours due to the orthodox theory.
According to this expression,
since the sensitivity is proportional to $\alpha_0^{-1/2}$ for $R_D=R_S$, 
one can see that 
calculated results are approximately scaled as the same power law.
However, for $Q_F=50$, the power law is no longer hold and for small tunneling 
resistances it is scaled as even  $\alpha_0^{1/2}$.
This behavior can be understood through a simple circuital analysis as in Ref. \cite{turin}
By replacing the tunneling barriers by resistors, it is found that
the best sensitivity is achieved for series resistance equal to $Q_F^2 R_0$ at which
input impedance of microwaves is matched to that of the tunneling barriers.
Thus, for $Q_F=6$ the matching condition occurs at $\alpha_0=1.5$ while for $Q_F=50$
it is expected to be $\alpha_0=0.02$.
As shown in Fig. \ref{senq0Fig}-$(b)$, this condition well agrees with the calculated result
for the small frequency even though for high frequencies it is slightly deviated
due to the non-adiabatic effects of microwaves.

In Fig. \ref{senq0Fig}-$(c)$ we show the dependence of the charge sensitivity on
tunneling resistances by omitting a calculational factor
one by one to emphasize its role.
According to results, considerations of self-consistent and non-adiabatic schemes
are found to be crucial for sensitivity calculations
while effects of the environment is relatively unimportant.
In addition, we find that symmetric geometry (dashed line) benefits the charge sensitivity
in the whole range of tunneling resistance.

\section{Summary}
In this work, we develop a formalism for 
a radio-frequency single-electron transistor 
taking into account electrical environment, higher-order co-tunneling,
and arbitrary time-dependent perturbations.
Assuming large charging energy, we use a two-charged-state model
in a metallic island and solved
the problem based on the Schwinger-Keldysh approach combined with
a generating functional method.
We calculate an approximated generating functional by summing diagrams in infinite order
and give exact expressions for current, charges in the island, and their noises
within the generating functional.
By defining generalized transport coefficients, we write the derived expressions
in terms of them, and show that
tunneling currents in time-dependence cases have a generalized form of
the well-known Landauer formula.

As application of our formalism, we examine tunneling currents, its noises, and the charge
sensitivity of rf-SETs by accounting for
a detailed tank circuit.
Firstly, effects of the electrical environments are found to be relatively small, as expected,
in cases of microwaves delivered via an coaxial cable with impedance $50\Omega$.
However, for a large quality factor and a large resonant frequency its effects become
large and cannot be ignored.
Secondly, effects of photon-assisted tunneling
are manifested to both enhanced responses and noises of rf SETs.
However, due to the larger enhancement of the noises, photon-assisted tunneling is not
helpful to the charge sensitivity.
As a consequence, with experimental parameters of Ref. \cite{schoel}, we obtain
the charge sensitivity of  $\delta q = 4.1\mu e/\sqrt{{\rm Hz}}$, which is larger than that
in the orthodox result, however, still much smaller
than the measure value of $47\mu e/\sqrt{\rm Hz}$.
Finally, we discuss the charge sensitivity depending on various sets of parameters.
Especially, we focus on its change as a function of tunneling resistance,
and find that the charge sensitivity for small quality factors
is scaled like $\alpha_0^{-1/2}$ as in the analytic result proposed by the previous work.
Whereas, for large quality factors the power law is no longer valid and it is
proportional to even $\alpha_0^{1/2}$ in the range of small tunneling resistance
to show optimal resistance for the best sensitivity.

\acknowledgments{
We thank Dr. H. J. Lee for the introduction to coherent-path-integral method and 
Dr. Y. S. Yu for providing his numerical results for comparison with ours
This work was supported by the Korean Ministry of Science and Technology
through the Creative Research Initiatives Program under Project No.
r16-1998-009-01001-0.}

%

\end{document}